\documentclass[fleqn,10pt,superscriptaddress]{wlscirep}
\usepackage{amsmath}
\usepackage{amssymb}
\usepackage{gensymb}
\usepackage{xargs}
\usepackage{dsfont}
\usepackage{color}
\usepackage{amsfonts}
\usepackage[normalem]{ulem}
\usepackage{upgreek}
\usepackage{bm}
\usepackage{graphicx}
\usepackage{hyperref}
\usepackage{color}

\usepackage{epstopdf, epsfig}
\usepackage[utf8]{inputenc}
\usepackage{upgreek}
\usepackage{soul}
\newcommand{\dif}{\textup d}

\allowdisplaybreaks[1]

\begin{document}

\title{High-flux neutron generation by laser-accelerated ions from single- and double-layer targets}

\author[1,2,3,*]{Vojtěch Horný}
\author[4]{Sophia N. Chen}
\author[2,3]{Xavier Davoine}
\author[1]{Vincent Lelasseux}
\author[2,3]{Laurent Gremillet}
\author[1]{Julien~Fuchs}

\affil[1]{LULI-CNRS, CEA, UPMC Univ Paris 06: Sorbonne Universit\'e, Ecole Polytechnique, Institut Polytechnique de Paris, F-91128 Palaiseau Cedex, France}
\affil[2]{CEA, DAM, DIF, F-91297 Arpajon, France}
\affil[3]{Universit\'e Paris-Saclay, CEA, LMCE, 91680 Bruy\`eres-le-Ch\^atel, France}
\affil[4]{Horia Hulubei National Institute for Physics and Nuclear Engineering, 30 Reactorului Street, RO-077125, Bucharest-Magurele, Romania}

\affil[*]{vojtech.horny@polytechnique.edu}

\date{\today}

\begin{abstract}
Contemporary ultraintense, short-pulse laser systems provide extremely compact setups for the production of high-flux neutron beams, such as those required for nondestructive probing of dense matter, research on neutron-induced damage in fusion devices or laboratory astrophysics studies. Here, by coupling particle-in-cell and Monte Carlo numerical simulations, we examine possible strategies to optimise neutron sources from ion-induced nuclear reactions using 1-PW, 20-fs-class laser systems. To improve the ion acceleration, the laser-irradiated targets are chosen to be ultrathin solid foils, either standing alone or preceded by a plasma layer of near-critical density to enhance the laser focusing. We compare the performance of these single- and double-layer targets, and determine their optimum parameters in terms of energy and angular spectra of the accelerated ions. These are then sent into a converter to generate neutrons via nuclear reactions on beryllium and lead nuclei. Overall, we identify configurations that result in neutron yields as high as $\sim 10^{10}\,\rm n\,sr^{-1}$ in $\sim 1$-cm-thick converters or instantaneous neutron fluxes above $10^{23}\,\rm n\,cm^{-2}\,s^{-1}$ at the backside of $\lesssim 100$-$\upmu$m-thick converters. Considering a realistic repetition rate of one laser shot per minute, the corresponding time-averaged neutron yields are predicted to reach values ($\gtrsim 10^7\,\rm n \,sr^{-1}\,s^{-1}$) well above the current experimental record, and this even with a mere thin foil as a primary target. A further increase in the time-averaged yield up to above $10^8\,\rm sr^{-1}\,s^{-1}$ is foreseen using double-layer targets. 
\end{abstract}

\maketitle


\section*{Introduction}

Neutrons have very distinctive characteristics compared to ions, electrons or x-rays, making them unique tools to investigate or modify the properties of materials. Their applicability extends far beyond nuclear physics, i.e., to fields as varied as material science~\cite{takenaka1999application, mor2015reconstruction, noguere2007non}, medical sciences~\cite{gray1943treatment}, transmutation of nuclear waste \cite{tajima2021spent}, laboratory astrophysics~\cite{chen2019extreme}, security \cite{brown1994application,sowerby2007recent}, biology~\cite{franchet1993radiolytic} or archaeology~\cite{constantinescu1999fast, gratuze1992ancient}. The tremendous progress seen in high-power laser technology within the last decades \cite{danson2019petawatt} enables one to design laser-based, bright neutron sources that could become complementary, and potentially more feasible, alternatives to larger-scale conventional facilities such as high-energy particle accelerators or nuclear fission reactors~\cite{Henderson_2013}. Moreover, laser-driven neutron sources are characterised by much higher density and shorter pulse duration \cite{pomerantz2014ultrashort, higginson2015temporal}, which could allow currently unforeseeable applications to be developed.
 
In the last decade, the generation of intense neutron pulses has been demonstrated in many laser experiments. The brightest sources reported so far have been obtained at the TRIDENT~\cite{roth2013bright} and PHELIX~\cite{kleinschmidt2018intense, gunther2022forward} facilities, both of which deliver laser pulses of 100~J energy and ps duration, yet with a low repetition rate of about one shot per hour. Both experiments were based on laser-acceleration of a deuteron beam from deuterated plastic targets via the target normal sheath acceleration (TNSA)~\cite{wilks2001, mora2003plasma} or breakout afterburner (BOA)~\cite{yin2011three, jung2013boa} mechanisms, and on subsequent $^9$Be($d,n$) nuclear reactions in a beryllium converter located a few mm behind the primary target. A variant of this method, whereby deuterons are driven by the laser radiation pressure in a near-critical-density plasma, has been recently investigated numerically under conditions relevant to the TRIDENT laser~\cite{huang2021high}.

The aim of this paper is, rather, to examine the neutron beams that could be produced using $(p,n)$ or $(d,n)$ reactions triggered by 1-PW-class, few-femtosecond laser systems, such as Apollon~\cite{burdonov2021characterization} or the upcoming ELI facilities in the Czech Republic~\cite{rus2017eli} and Romania~\cite{gales2018extreme, radier_HPLSE_2022}, and which are as well commercially available. These systems usually operate at the frequency of one shot per minute. This limitation, however, is set by the internal nuclear safety rules of those institutions rather than by technological constraints and, in fact, such systems could in principle run at $10\,\rm Hz$~\cite{bayramian2008high}. Provided that their temporal contrast can be much improved over present performance, through, e.g., the use of plasma mirrors~\cite{levy2007double}, we can surmise that those lasers, which already surpass the $10^{21}\,\rm W\,cm^{-2}$ intensity level~\cite{burdonov2021characterization, Yoon_Optica_2021}, can boost the ion acceleration to the 100~MeV range.
One appealing prospect towards high-efficiency, laser-based neutron sources would be to couple such energetic ion beams with heavy-ion converters to approach the spallation regime of neutron generation, characterised by high neutron multiplicity~\cite{filges2009handbook, martinez2022numerical}.

Many schemes have been proposed to enhance the energy of the ions produced by ultraintense laser pulses, and hence their neutron production efficiency. The most straightforward, albeit challenging as regards the laser contrast, makes use of simple nanometer-thick solid foil targets~\cite{esirkepov2006laser, brantov2015ion}. This bears the promise of accelerating ions in a hybrid regime, governed by radiation pressure~\cite{henig2009rpa} or light-sail acceleration~\cite{esirkepov2004highly, klimo2008monoenergetic} at early times, and then followed by another mechanism (e.g. TNSA, breakout afterburner, or Coulomb explosion \cite{qiao2012dominance, gong2020proton}) that boosts the energy of the ions. The energy gain in this scheme is predicted to be increased by the onset of relativistic transparency during the laser-plasma interaction \cite{esirkepov2006laser, Ferri2020}. 

Another route, which has attracted great interest in the past decade, is to exploit double-layer targets (DLTs). In these, a near-critical density (NCD) plasma layer, having thickness of tens of $\upmu$m and serving as a lens to focus the laser~\cite{esarey1997self, Wang2011}, is attached to an overdense/solid, plastic or metal thin foil (with thickness from tens of nm to a few $\upmu$m). Due to relativistic self-focusing in the NCD layer, the intensity of the laser pulse can rise multiple times over its initial value \cite{pazzaglia2020theoretical}. Meanwhile, a significant fraction of its energy can be converted into relativistic electrons, either via direct laser-electron interaction \cite{Quesnel1998, Salamin2002}, resonant-type coupling of the laser and plasma fields \cite{Pukhov1999, arefiev2016beyond}, or strongly nonlinear plasma wakefields \cite{Debayle2017}. The boosted hot-electron generation and laser focusing achieved in DLTs can both contribute to increase the ion energy, by strengthening either the accelerating sheath fields at the target surfaces or the laser radiation pressure. Predicted numerically \cite{nakamura2010foam, sgattoni2012laser, wang2013efficient, passoni2016toward, levy2020enhanced}, the improved performance of DLTs has been confirmed in a number of experiments \cite{yogo2008laser, passoni2014energetic, bin2015ion, passoni2016toward, bin2018enhanced, ma2019laser}. Practically, DLTs can be manufactured employing available technologies, e.g., by deposition of nanostructures \cite{bin2015ion, bin2018enhanced, passoni2016toward, ma2019laser} or foams \cite{prencipe2016development, wang2021fabrication} on the irradiated side of the foil.

In this paper, using particle-in-cell (PIC) and Monte Carlo (MC) numerical simulations, we characterise in detail the neutron beams resulting from the interaction of 1-PW, 20-fs laser pulses (modeling the pulses currently produced at Apollon~\cite{burdonov2021characterization}) with single- (SLTs) and double-layer (DLTs) targets. By performing scans in the laser focusing, primary target and converter parameters, we identify the conditions maximizing the yield, areal density and peak flux of the neutron source. One major finding is that while the maximum number of neutrons ($\sim 2.8\times 10^{10}$) is obtained with DLTs combined with Pb converters, the maximum neutron yield per solid angle ($\sim 1.6\times 10^{10}\,\rm n\,sr^{-1}$) is achieved using a deuterated DLT and a Be converter. In addition, very thin ($<100\,\rm \upmu m$) converters are predicted to generate peak neutron fluxes in excess of $\sim 10^{23}\,\rm cm^{-2}\,s^{-1}$, which opens a path towards applications in laboratory astrophysics~\cite{chen2019extreme}.

\begin{figure*}
    \centering
    \includegraphics[scale=1.0]{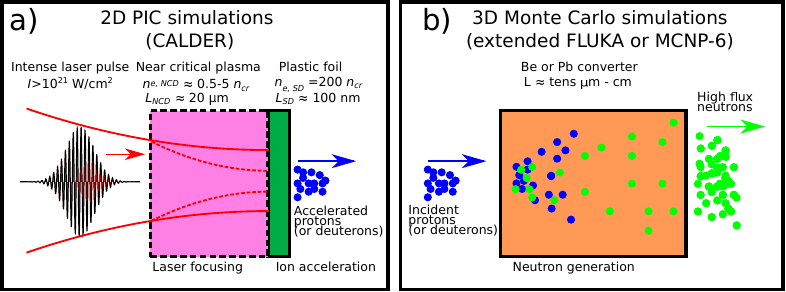}
    \caption{\textbf{Numerical methodology.} (a) In the first stage, 2D PIC simulations are performed, using the \textsc{calder} code, to optimise proton (or deuteron) acceleration with either single- or double-layer targets (see text). (b) The accelerated protons (or deuterons) are then sent to the 3D Monte Carlo \textsc{fluka} or \textsc{mcnp-6}
    code to compute the neutron generation in a secondary beryllium or lead converter. 
       }
    \label{fig:scheme}
\end{figure*}

\section*{Results}

\subsection*{Simulation setup}

Our methodology is outlined in Fig.~\ref{fig:scheme}. We first use the particle-in-cell (PIC) \textsc{calder} code (see Methods) to simulate in 2D3V (two-dimensional in space, three-dimensional in momentum space) geometry, the laser-matter interaction and proton (or deuteron) acceleration from either SLTs or DLTs. In a second step, the accelerated ions are transferred to a three-dimensional (3D) Monte Carlo code which describes neutron generation in a secondary converter target (see Methods).

The laser parameters in the PIC simulations are chosen to match those already accessible at the Apollon facility~\cite{burdonov2021characterization}, but assuming improved temporal contrast conditions -- such as those expected from fielding a plasma mirror system \cite{levy2007double} -- in order to enable efficient interaction of the laser pulse with nanometer-scale foils, as investigated in the following. The laser is modeled as a Gaussian pulse of central wavelength $\lambda_{\rm L} = 0.8\,\rm \upmu m$ and $\tau_{\rm L} = 20\,\rm fs$ full-width-at-half-maximum (FWHM) duration. It is focused to a $D_{\rm L} = 5\,\rm \upmu m$ FWHM spot at the front side of the target (i.e. the front side of the NCD plasma layer in the case of a DLT). Its peak intensity is $I_0 = 2\times 10^{21}\,\rm W\,cm^{-2}$, corresponding to a dimensionless field strength $a_0 = eE_0/m_e c\omega_{\rm L} = 30.6$ ($E_0$ is the laser electric field, $c$ the light speed, $e$ the elementary charge, $m_e$ the electron mass, and $\omega_{\rm L} = 2\pi c/\lambda_{\rm L}$ the laser frequency). PIC simulations are performed in the $x-y$ plane. The laser pulse is linearly polarised along $y$ and propagates in the $+x$ direction. In a realistic 3D geometry, the laser pulse energy and power would be of 22~J and 1~PW, close to the current Apollon parameters.

As detailed in Table~\ref{tab:simpar}, the simulated DLTs consist of a submicron-thick, fully ionised plastic CH$_2$ (or CD$_2$) foil of solid density, preceded by a fully ionised carbon (C$^{6+}$) NCD plasma layer of varying density ($n_{e,\rm NCD}$) and length ($l_{\rm NCD}$) . The NCD parameters optimizing proton acceleration from DLTs have been investigated both numerically and analytically by Pazzaglia \emph{et al.} ~\cite{pazzaglia2020theoretical}. In that study, the maximum proton cutoff energies were attained for NCD layers of thickness close to the relativistic self-focusing length, and over a limited density range. The following approximate formulas were obtained for the optimal length and density of the NCD layer [see Eqs.~(23) and (24) in Ref.~\citen{pazzaglia2020theoretical}]:
\begin{align}
  l_{\rm NCD} &= 0.88\,\frac{D_{\rm L}^2/\lambda_{\rm L}}{(\tau_{\rm L}c/\lambda_{\rm L})^{1/3}} \,, \label{eq:l_ncd} \\
  n_{e, \rm NCD} &= 0.91\,\gamma_0 n_{\rm cr} \frac{\lambda_{\rm L}^2}{D_{\rm L}^2} \left( \tau_{\rm L} c/\lambda_{\rm L} \right)^{2/3} \,, \label{eq:ne_ncd}
\end{align}
where $\gamma_0 = \sqrt{1+a_0^2/2}$ is the mean Lorentz factor of the laser-driven electrons and $n_{\rm cr} [\rm cm^{-3}] = 1.1 \times 10^{21}\,\lambda_L^{-2} [\rm \upmu m]$ is the nonrelativistic electron critical density. For our parameters, $\gamma_0 = 21.7$, $n_{e,\rm NCD} = 1.93\,n_{\rm cr}$ and $l_{\rm NCD} = 14.1\,\rm \upmu m$. These values will provide reference conditions for our PIC simulations.

\subsection*{Laser self-focusing in the near-critical plasma}

\begin{figure*}
    \centering
    \includegraphics[scale=1.0]{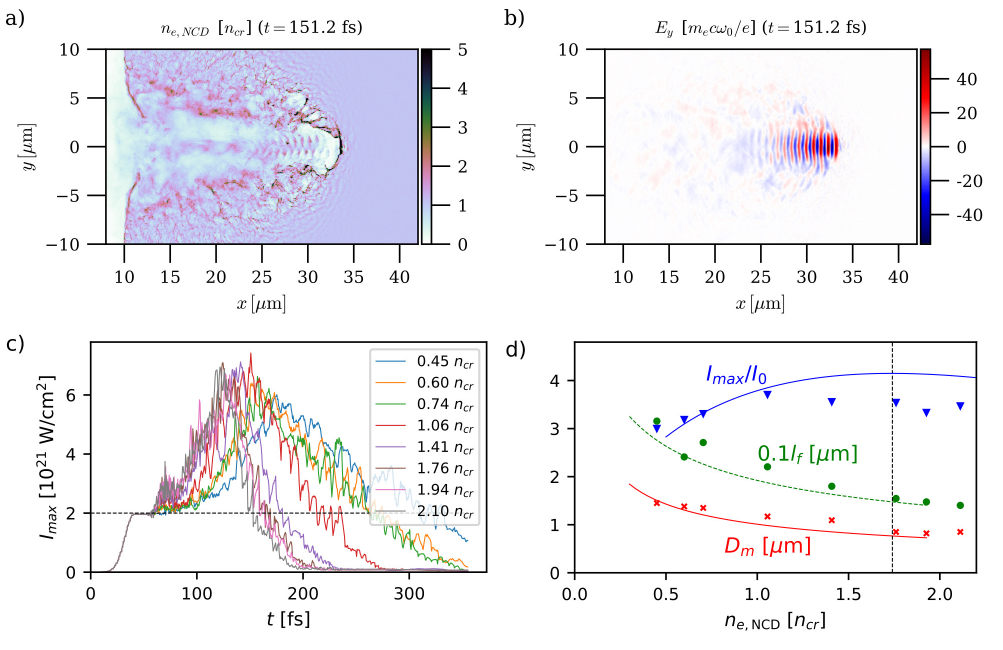}
    \caption{\textbf{Relativistic self-focusing of a $\mathbf{2\times 10^{21}}\,\rm \mathbf{W\,cm^{-2}}$, 20~fs duration laser pulse in near-critical plasmas.} Spatial distributions of (a) the electron density $n_e$ (saturated colormap) and (b) $E_y$ electric field component for an initial plasma electron density $n_{e, \rm NCD} = 1.06\,n_{\rm cr}$ at the simulation time $t=136.9\,\rm fs$, i.e., 99.3~fs after the laser pulse maximum entered the simulation box. (c) Time evolution of the instantaneous peak intensity of the laser pulse during its propagation through the plasma for eight different values of $n_{e,\rm NCD}$. The horizontal dashed line indicates the peak laser intensity in vacuum, $I_0 = 2\times 10^{21}\,\rm W\,cm ^{-2}$. (d) Dependence of the maximum intensity increase factor $I_{\rm max}/I_0$ (blue triangles and solid line), focusing distance $l_f$ (green circles and dashed line), and FWHM spot size $D_m$ (red crosses and solid line) of the laser pulse on the initial plasma density. The symbols represent the PIC simulation results while the curves correspond to the formulas given by Pazzaglia \emph{et al.}~\cite{pazzaglia2020theoretical}. The vertical dashed line indicates the plasma density yielding the best focusing according to analytical theory.
    }
    \label{fig:fig2}
\end{figure*}

To start with, we inspect the process of laser self-focusing in the NCD plasma layer. To illustrate it, we consider the case of $n_{e, \rm NCD}=1.06 n_c$. Figures~\ref{fig:fig2}(a,b) depict the spatial distributions of, respectively, the electron density ($n_e$) and transverse electric field ($E_y$) at a time ($t=151\,\rm fs$) when the laser has propagated about $21\,\rm \upmu m$ in the plasma. An electron density channel has then formed along the laser path: the laser pulse is concentrated into a $D_m = 1.1\,\rm \upmu m$ FWHM spot at the head of the channel, where its peak intensity reaches $\sim 7.4\times 10^{21}\,\rm W\,cm^{-2}$ -- an increase by a factor of $\sim 3.7$ over its incident value. Inside the channel, we find (not shown) that a large fraction of the plasma electrons are accelerated by the laser wave and/or wakefields \cite{arefiev2016beyond, Debayle2017, ma2019laser, levy2020enhanced} to Lorentz factors as high as $\gamma \simeq 300-400$, largely exceeding the standard ponderomotive scaling, $\langle \gamma \rangle \simeq \gamma_0$.

Figure~\ref{fig:fig2}(c) plots the temporal evolution of the maximum intensity $I_{\rm max} = \frac{c\varepsilon_0}{2}(E_x^2+E_y^2+E_z^2)$ achieved within the simulation box for eight different initial electron densities in the NCD plasma, in the range $0.45 \le n_{e,\rm NCD}/n_{\rm cr} \le 2.10$. All cases lead to a significant (by a factor of $\gtrsim 3$), and quite comparable, enhancement of the intensity compared to its value in vacuum ($I_0 = 2\times 10^{21}\,\rm W\,cm^{-2}$, shown as a dashed line). The maximum intensity is reached after an interaction time of $\sim 120-180\,\rm fs$, decreasing with density. The $n_{e,\rm NCD}= 1.06\,n_{\rm cr}$ case, corresponding to Figs.~\ref{fig:fig2}(a,b), is found to yield, though by only a very small margin, the highest intensity amplification.  

Figure~\ref{fig:fig2}(d) shows the variations with initial plasma density $n_{e,\rm NCD}$ in the laser intensity amplification factor $I_{\rm max}/I_0$ as well as in the associated focal spot $D_m$ and length $l_f$. For each value of $n_{e,\rm NCD}$, those quantities are recorded at the time and location corresponding to the peak instantaneous laser intensity. For comparison, the theoretical estimates of $D_m$ and $l_f$, based on the thin-lens approximation, and given by Eqs.~(2) and (4) of Ref.~\citen{pazzaglia2020theoretical} are displayed as solid red and dashed green curves, respectively. Also overlaid (blue solid line) is the intensity enhancement factor $I_{\rm max}/I_0$, as obtained by solving numerically Eqs.~(7) and (8) of Ref.~\citen{pazzaglia2020theoretical}. The dashed line indicates the density ($n_{e,\,\rm NCD} = 1.74\,n_{\rm cr}$) that maximises the intensity enhancement ($I_{\rm max}/I_0 = 4.1$) according to that model.

Overall, the stronger lensing effect of the plasma at larger density is clearly demonstrated, and the theoretical predictions 
match well with the simulation data.
Consistent with Fig.~\ref{fig:fig2}(c), the simulated intensity enhancement is found to weakly vary (by $\sim 3.0-3.7$) in the density range investigated, $0.45 \le n_{e,\rm NCD}/n_{\rm cr} \le 2.1$. While $I_{\rm max}/I_0$ slightly decreases (from $\sim 3.7$ to $\sim 3.3$) when the plasma density is raised from $n_{e,\rm NCD} = 1.06\,n_{\rm cr}$ to $2.1\,n_{\rm cr}$, it drops relatively more abruptly when the plasma density is decreased from the optimal density of $1.06\,n_{\rm cr}$ (down to $I_{\rm max}/I_0 \simeq 3.0$ at $n_{e,\rm NCD} = 0.45$).

It is worth noting that the observed variation in $I_{\rm max}$ is weaker than that one would naively infer assuming laser energy conservation (and hence $I_{\rm max} \propto D_m^{-2}$) from the concomitant variation in spot size. As the latter quantity shrinks from $D_m \simeq 1.5\,\rm \upmu m$ to $\simeq 0.8\,\rm \upmu m$ as $n_{e,\rm NCD}$ increases, one would then expect $I_{\rm max}$ to increase by a factor $\sim (1.8/0.8)^2 \sim 3.5$, at odds with the simulation data. This discrepancy points to the strong dissipation undergone by the laser pulse during its propagation in the NCD plasma. Actually, in their model Pazzaglia \emph{et al.} made an attempt at taking into account laser dissipation into hot electron generation. While the simulations and the model yield comparable maximum values of the intensity increase ($I_{\rm max}/I_0 \simeq 3.7$ vs $\simeq 4.1$), the corresponding optimal plasma densities are significantly lower in the simulations ($n_{e,\rm NCD} = 1.06\,n_{\rm cr}$) than theoretically predicted ($n_{e,\rm NCD} \simeq 1.5-2\,n_{\rm cr}$).
These results suggest that, for our simulation parameters, the actual dissipation is higher than described analytically, possibly due to fast electrons being energized much beyond the ponderomotive level considered in the model~\cite{pazzaglia2020theoretical}. For instance, the excitation of strong plasma waves could provide an efficient additional source of energy depletion of the laser pulse \cite{Shadwick_PoP_2009, Debayle2017}.

We conclude this part by mentioning that in an actual 3D configuration, relativistic laser self-focusing is expected to lead to even stronger intensification of the laser pulse, as previously shown in Ref.~\citen{pazzaglia2020theoretical}.

\begin{table*}
\begin{center}
\begin{tabular}{|c|c|c|c|c|c|c|c|c|c|c|}
\hline
\# & $l_{\rm NCD}$ [$\upmu$m] & $l_{\rm SD}$ [nm] & $n_{e,\rm NCD}$ [$n_{\rm cr}$] & SD material & $N_p$ or $N_d$ & $E_{\rm cutoff}$ [MeV] \\
\hline
1  & -     & 64  & -    & CH$_2$ & $1.29\times 10^{12}$ & 128 \\
2  & -     & 115 & - & CH$_2$ &  $1.36\times 10^{12}$ & 94 \\
\hline
3  & 20.46 & 115 &so1.06 & CH$_2$ &  $1.46\times 10^{12}$ & 239  \\
4  & 25.19 & 115 & 0.74  & CH$_2$ &  $9.86\times 10^{11}$ & 241 \\
5  & 17.40 & 115 & 1.06  & CH$_2$ &  $1.32\times 10^{12}$ & 218 \\
\hline
3b  & 20.46 & 115 &  1.06 & CD$_2$ &  $5.95\times 10^{11}$ & 224 \\
\hline
\end{tabular}
\end{center}

\caption{\textbf{Parameters
of the proton acceleration simulations.} The physical parameters of the single- and double-layer simulations are: Thickness of the near-critical density (NCD) carbon plasma layer ($l_{\rm NCD}$), thickness of the solid-density (SD) CH$_2$ (or CD$_2$) layer ($l_{\rm SD}$), electron density of the NCD layer ($n_{e,\rm NCD}$), electron density of the SD layer $n_{e, \rm SD} = 200\,n_{\rm cr}$, laser spot size at the front of the NCD layer (if present, otherwise at the front of the SD layer) $D_L = 5\,\rm \upmu m$, dimensionless laser field strength $a_0 =30.6$, total number of accelerated protons or deuterons with energy above 1~MeV ($N_p$, $N_d$), ion cutoff energy ($E_{\rm cutoff}$).
}
\label{tab:simpar}
\end{table*}

\subsection*{Proton acceleration from single- and double-layer targets}

We now assess the performance of SLT- or DLT-based ion acceleration setups in generating intense neutron fluxes from secondary converter targets. Table~\ref{tab:simpar} summarises the parameters of the six SLT and DLT simulations we have carried out. Runs \#1 and \#2 represent SLT configurations. The thickness of the CH$_2$ foil in run~\#1 is that predicted to optimise the proton cutoff energy under our irradiation conditions according to Refs.~\cite{brantov2007ion, brantov2015ion} [see Eq.~\eqref{eq:brantov} in Methods]. The 115-nm-thick CH$_2$ foil in run~\#2 is optimised for the maximum intensity of $7.8 \times 10^{21}\,\rm W\,cm^{-2}$ achieved during the laser propagation in the NCD plasma. This foil is about twice thicker than in run~\#1 but still partly transparent; a higher number of protons should then be produced but at lower energies. Runs \#3-5 use DLTs with NCD layers of different densities or lengths, coated on the same CH$_2$ foil as in run \#2 to account for plasma lensing. In runs \#3 and \#4, the NCD layer thickness is set to the laser focusing length in order to achieve the highest possible laser intensity, and thus to maximise proton energies. The combinations of the $n_{e,\rm NCD}$ and $L_{\rm NCD}$ values in runs \#3 and \#4 are those yielding the strongest laser intensities [see Fig.~\ref{fig:fig2}(c)]. Run \#5 is similar to run \#3 but uses a shorter $L_{\rm NCD}$ to examine how the final proton and neutron beams depend on the NCD layer length. Finally, run \#3b has the same parameters as run \#3 but protons are here fully replaced with deuterons (i.e. the solid foil is made of CD$_2$). We acknowledge that our comparison of DLTs with CH$_2$ and CD$_2$ solid foils is somewhat idealised because, in reality, (i) there would be proton-rich contaminant layers on either side of the foil, whatever its composition and (ii) deuterated targets used for laser-plasma acceleration often contain a $\gtrsim 1\,\%$ fraction of hydrogen within their bulk \cite{roth2013bright}.

In all PIC simulations, a virtual detector is placed at a distance of $26.2\,\rm \upmu m$ behind the rear side of the target. When an accelerated proton crosses this ``plane'', its position, momentum and time of arrival are stored.  A relatively close detector position is chosen in order to interrupt the acceleration process. This is a common practice because 2D~PIC simulations are known to overestimate the final proton energy \cite{babaei2017rise}. Our choice of distance relies on a comparison between 2D and 3D PIC simulations~\cite{brantov2015ion}.

\begin{figure*}
    \centering
    \includegraphics[scale=1.0]{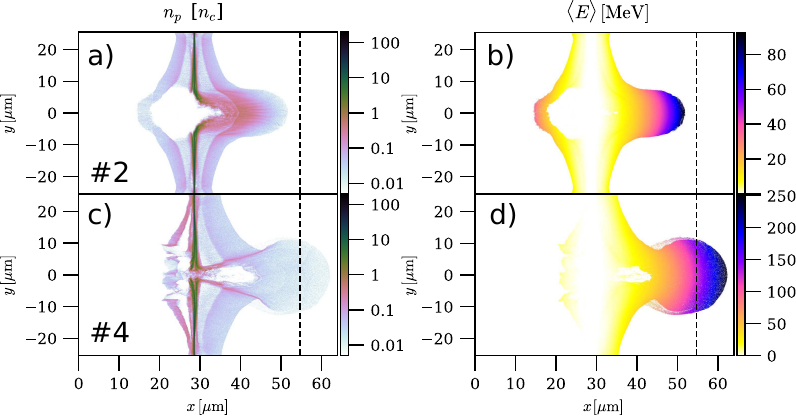}
    \caption{\textbf{Visualisation of the proton acceleration process. } Number density (a,c) and  local average energy (b,d) of protons in runs \#2 (a,b) and \#4 (c,d), 207~fs after the laser pulse maximum has reached the solid foil. The dashed black line marks the position of the virtual detector where the quantities shown in Fig.~\ref{fig:protspec} are measured. }
    \label{fig:densnap}
\end{figure*}

Figure~\ref{fig:densnap} displays the spatial distributions of the proton number density (a,c) and average energy (b,d), 207~fs after the laser pulse maximum has hit the solid foil. The top and bottom rows correspond to SLT run \#2 and DLT run \#4, respectively. In both cases, the target protons are fully evacuated from the laser spot region and preferentially accelerated in the forward direction due to the combined effects of RPA and TNSA [see Figs.~\ref{fig:densnap}(a, c)]. Compared to the SLT case, the tighter laser focusing achieved in the DLT translates into a more energetic (by a factor of $\sim 3$), yet more divergent, proton beam originating from a narrower source [cf. Figs.~\ref{fig:densnap}(b, d)].

\begin{figure}
    \centering
    \includegraphics[scale=1.0]{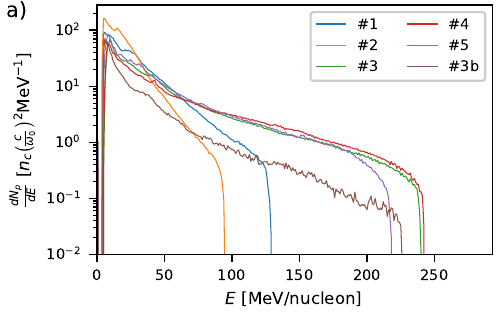}
    \includegraphics[scale=1.0]{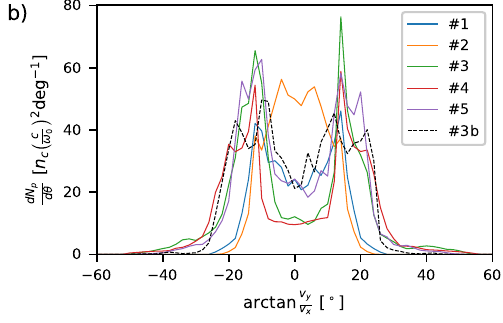}
    \caption{\textbf{Properties of accelerated protons from 2D PIC simulations. } (a) Energy and (b) angular spectra of protons tracked at the virtual detector, $26.2\,\rm \upmu m$ behind the target [marked by the vertical dashed line in all the panels of Fig.~\ref{fig:densnap}].
    }
    \label{fig:protspec}
\end{figure}

Figure~\ref{fig:protspec}(a) compares the energy spectra of the protons (or deuterons) that reached the detector plane [marked by vertical dashed lines in Figure~\ref{fig:densnap} for the six configurations considered. The absolute maximum proton energy ($\sim 240\,\rm MeV$) is recorded in DLT runs  \#3 and \#4, as expected from the associated laser intensification [see Fig.~\ref{fig:fig2}]. By comparison, the SLT cases yield significantly lower proton energies: $\sim 130\,\rm MeV$ in the optimised run \#1 and $\sim 95\,\rm MeV$ for the about twice thicker foil of run \#2.

As already mentioned, the flip side of the fastest protons produced in DLTs is their increased angular spread, due to strong focusing and partial transmission of the laser pulse through the foil. 
Actually, their angular spectra plotted in Fig.~\ref{fig:protspec}(b), turn out to be double-peaked at $\sim \pm 15-20^\circ$ and to extend up to angles as large as $\sim \pm 40^\circ$.
By contrast, SLTs can provide more collimated protons and also in greater numbers (see Table). Run \#2 indeed yields a proton angular spectrum peaking on axis and with an FWHM spread of $\sim 12.5^\circ$. 

Deuterons in DLT run \#3b attain a similar, albeit slightly lower, cutoff energy ($\sim 225\,\rm MeV$) than protons in a similar setup (run \#4), but their number is about $2-4\times$ lower in the high-energy ($>100\,\rm MeV$) part of the spectrum. Their angular distribution also peaks off axis but with a less pronounced on-axis minimum than in the proton cases. We note that while the absence of surface protons in our simulations is expected to favour, to some extent, the deuteron acceleration, it was shown experimentally \cite{roth2013bright} that a large majority of the ions accelerated from 300~nm CD$_2$ foils were deuterons despite a deuterisation level only as high as $90\,\%$.

\subsection*{Proton transport and neutron generation through the converter}

\begin{figure}
    \centering
    \includegraphics[scale=1.0]{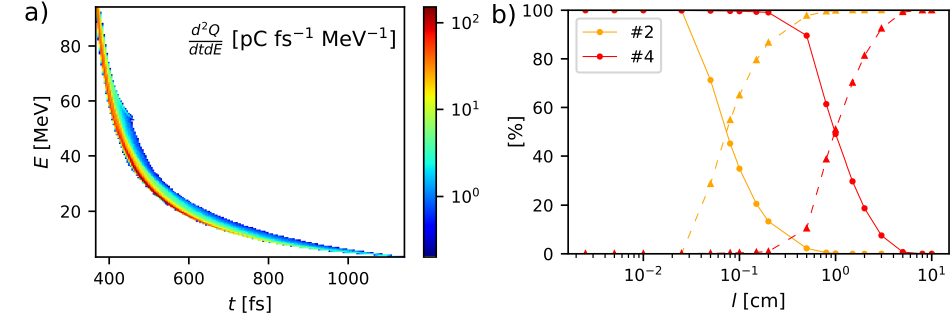}
    \caption{\textbf{Proton transport through a Pb converter.} (a) Time-resolved energy spectrum of the protons crossing the virtual detector as recorded in SLT run \#2 (see Table~\ref{tab:simpar}). 
    (b) Dashed lines: dissipated fraction of proton beam energy due to inelastic collisions as a function of the Pb converter length $l$, in SLT case \#2 (orange) and DLT case \#4 (red). Solid lines: transmitted fraction of proton beam energy in SLT case \#2 (orange) and DLT case \#4 (red).}
    \label{fig:spectrogram}
\end{figure}

The accelerated protons (or deuterons) recorded by the virtual detector in the PIC simulations are used as inputs to the Monte Carlo simulations of their transport through the neutron converter target. The latter, located $26.2\,\rm \upmu m$ away from the laser target, consists of a lead ($^{208}$Pb) or beryllium ($^9$Be) cylinder of 3-cm radius and varying length ($l$). This procedure is illustrated in Fig.~\ref{fig:spectrogram}. Figure~\ref{fig:spectrogram}(a) displays the time-resolved energy spectrum of the outgoing protons in SLT case \#2: it is characterised by a maximum energy of $\simeq 95\,\rm MeV$ (i.e., the lowest cutoff energy among our simulations), and a root-mean-square pulse duration of $\simeq 190\,\rm fs$ (the time interval between the incidence of the first and last protons with energy above 1~MeV is 770~fs). The dashed orange curve in Fig.~\ref{fig:spectrogram}(b) plots the fraction of energy lost (via inelastic collisions) by those protons in the Pb converter, as a function of its length. The dashed blue curve plots the same quantity for the proton beam generated in DLT case \#4, associated with the highest ($\sim 250\,\rm MeV$) cutoff energy. The orange and blue solid lines represent, respectively for cases \#2 and \#4, the (complementary) fraction of proton energy transmitted across the converter's backside. For case \#2 (resp. case \#4), the beam energy dissipation remains negligible ($\leq 1\,\% $) in Pb converters thinner than $\simeq 250\,\rm \upmu m$ (resp. $\simeq 2\,\rm mm$), while it is almost complete ($\geq 90\,\%$) for $l \geq 3\,\rm mm$ (resp. $l \geq 3\,\rm cm$). 

\begin{figure}
    \centering
    \includegraphics[scale=1.0]{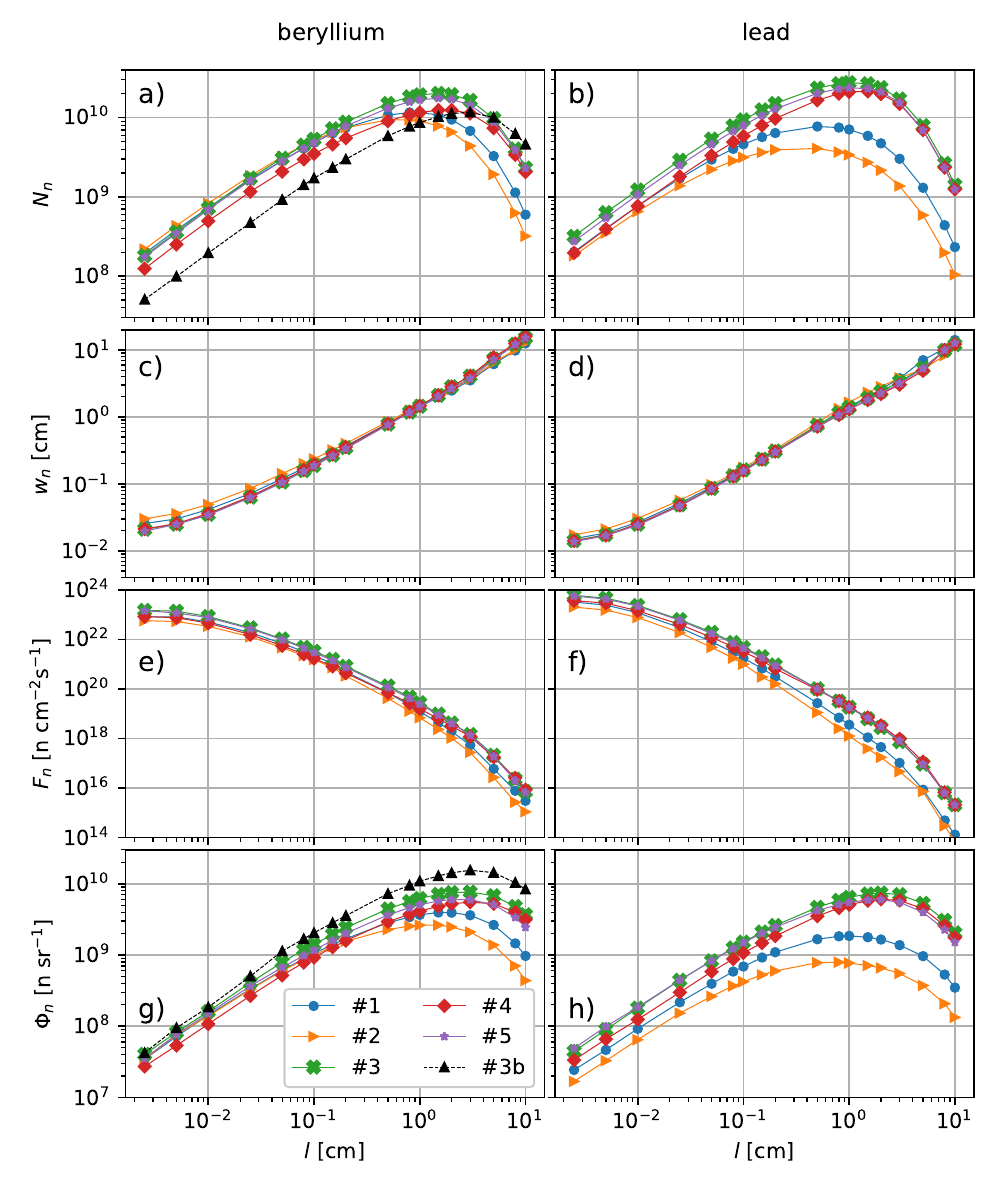}
    \caption{\textbf{Properties of the neutron source as a function of the converter length. }(a,b) Total yield, (c,d) transverse beam size, (e,f) peak instantaneous flux and (g,h) on-axis yield per unit solid angle. The first three quantities are measured at the backside of the converter while the latter is recorded on axis, 20~cm behind the converter. The left and right columns correspond to the Be and Pb converters, respectively. The labels of the different curves are as listed in Table~\ref{tab:simpar}.
    }
    \label{fig:nsource}
\end{figure}

Figure~\ref{fig:nsource} details the variations with the converter length of the properties of the neutron sources generated in beryllium (left column) and lead (right column), for the ion-acceleration setups listed in Table~\ref{tab:simpar}. Only the neutrons leaving the backside of the converter are recorded (i.e., those escaping from the front and lateral sides are excluded). Figures~\ref{fig:nsource}(a,b) show that the highest absolute neutron yield, $N_n \simeq 2.8 \times 10^{10}$, is achieved in a 1-cm-thick Pb converter exposed to the protons generated in DLT run \#3 -- the setup generating the greatest number of fast protons. Other DLT runs \#4 and \#5 give very close results. The higher energy and number of protons accelerated in DLTs translate into a $\sim 40\,\%$  greater total yield in Pb than in Be. The enhanced performance of Pb converters coupled with DLTs can be partly ascribed to a larger $^{208}$Pb$(p,n)$ cross section, which reaches $\sim 1\,\rm b$ at a $\sim 13\,\rm MeV$ proton energy and keeps on rising at higher energy, exceeding $\sim 10\,\rm b$ at $\sim 100\,\rm MeV$ \cite{Brown_NDS_2018}. This is unlike the $^9$Be$(p,n)$ cross section which peaks around $\sim 5\,\rm MeV$ with a value of $\sim 0.2\,\rm b$, and drops at higher energy, falling below $\sim 10\,\rm mb$ above $\sim 100\,\rm MeV$ \cite{Soppera_OECD_2020}.

When using SLTs, by contrast, the total neutron yield is in general larger in Be, by tens of per cent in run \#1 and by up to three times in run \#2 the one giving the slowest protons). The maximum yield ($N_n \simeq 1.2 \times 10^{10}$) is then recorded in a 0.8-cm-thick Be converter using proton source \#1.

For a given ion beam, the $N_n$ vs. $l$ curve peaks around the beam penetration distance in the converter. The decreasing trend at larger $l$ mainly originates from neutrons escaping from the lateral sides of the converter; such side losses become significant when the transverse size of the neutron beam, $w_n$ [plotted vs. $l$ in Figs.~\ref{fig:nsource}(c,d)], approaches the converter's diameter. The beam size is obtained by fitting to a Gaussian the neutron fluence profile at the backside of the converter. Within the range of converter lengths considered, $w_n$ is approximately equal to $l$, except in very thin converters ($l \lesssim 100\,\rm \upmu m$), where it is mainly set by the initial transverse proton beam size and divergence to a lower value of $\sim 150\,\rm \upmu m$. The impact of the converter's shape and size on the neutron yield in accelerator-based spallation neutron sources is discussed in Ref.~\citen{filges2009handbook}.

In Pb converters of thickness $l < 1\,\rm cm$, the total neutron yield is maximised in DLT runs \#3 and \#5, while the three DLT setups give similar results for $l > 1\,\rm cm$. For thin Be converters ($l \lesssim 2\,\rm mm$), the neutron yield is maximised, to within small differences, in SLT runs \#1 and \#2 and DLT runs \#3 and \#5. For $0.2 < l \lesssim 3\,\rm cm$, runs \#3 and \#5 perform the best and produce the maximum absolute value $N_n \simeq 2\times 10^{10}$ around $l\simeq 2\,\rm cm$. In thicker Be converters $N_n$ steadily decreases and is optimised in the deuteron-based run \#3b.

The peak neutron flux, $F_n$, plotted in Figs.~\ref{fig:nsource}(e,f) is a relevant parameter for certain purposes such as laboratory studies on $r-$process nucleosynthesis \cite{chen2019extreme, tain2015enhanced, tonchev2017capture}.  It is calculated as
\begin{equation}
F_n = \frac{4}{\pi w_n^2}\frac{\dif N_n}{\dif t} \,,    
\end{equation}
where $\dif N_n/\dif t$ is obtained from our extended \textsc{fluka} routine at the converter's backside. It is seen that neutron fluxes exceeding $10^{23}\,\rm n\, cm^{-2}\,s^{-1}$ can be attained using Be or Pb converters a few $10\,\rm \upmu m$ thick only, in which the neutron source is the narrowest and the shortest. In an actual experiment, it would then be important to place the converter as close as possible to the ion-generating target, while not hindering the ion acceleration process, as is the case in our simulation setup. Note that target assemblies consisting of two solid foils separated by tens of microns have already been employed in laser-based ion acceleration experiments, notably in order to tune the cutoff energy of TNSA ions~\cite{Chen_PoP_2014}.

The record peak flux ($\sim 6\times 10^{23}\,\rm n\,cm^{-2}\,s^{-1}$) is achieved, though by a very short margin, in DLT case \#3 with a $\sim 25$-$\upmu$m-thick Pb converter. Other DLT cases yield very similar results while the SLT cases perform only slightly less well. For $\sim 1 \,\rm cm$ converter lengths maximizing the neutron yield, the peak flux drops to $10^{18}-10^{19}\,\rm n\, cm^{-1}\,s^{-1}$, with a larger difference between SLT and DLT data in Pb than in Be.

To assess the dependence of the neutron flux on the distance ($d$) between the laser target and the converter, we have performed two additional Monte Carlo simulations using the ion source from DLT case \#3 but with a Pb converter located $1\,\rm m m$ away from the laser target. The peak flux at the backside of a thin ($25\,\rm \upmu m$) converter is then reduced to $F_n \simeq 9.1\times 10^{22}\,\rm n \,cm^{-2}\,s^ {-1}$, i.e., $\sim 15\,\%$ of the value achieved in our baseline configuration ($d=26\,\rm \upmu m$). For a 2-cm-thick converter, we obtain $F_n \simeq 2.1\times 10^{18}\,\rm n \,cm^{-2}\,s^ {-1}$, i.e., $\sim 75\,\%$ of the value attained previously.

Figures~\ref{fig:nsource}(g,h) display the neutron yield per unit solid angle, $\Phi_n$, i.e., the important quantity for most applications. $\Phi_n$ is measured on axis 20~cm away from the rear side of the converter. When using protons, $\Phi_n$ is maximised with DLTs. Specifically, one finds that $\Phi_n$ reaches similar peak values ($\simeq 8\times 10^9\,\rm n\,sr^{-1}$) in $\sim 2-3$-cm-thick Be or Pb converters exposed to proton beams from DLT case \#3. It is worth noting that the lower-energy protons from both SLT cases are only slightly less efficient in Be ($\Phi_n \simeq 2.5-4\times 10^9\,\rm n\,sr^{-1}$) than DLT protons but perform significantly poorer in Pb ($\Phi_n \simeq 0.8-2\times 10^9\,\rm n\,sr^{-1}$). A major finding, however, is that for Apollon-class laser parameters, the highest absolute neutron yield per solid angle ($\Phi_n \simeq 1.6\times 10^{10}\,\rm n\,sr^{-1}$) is achieved by deuterons from DLT run \#3b in a 3-cm-thick Be converter [see Fig.~\ref{fig:nsource}(g)]. This twofold increase in $\Phi_n$ may seem surprising as deuterons produce an approximately three times lower neutron yield than DLT-accelerated protons in Pb. The origin of this result is the pronounced forward directionality of deuteron breakup neutrons for deuteron energies above $\sim 2.2\,\rm MeV$ \cite{weaver_NSE_1972, avrigeanu2010deuteron}.

\begin{figure}
    \centering
    \includegraphics[scale=1.0]{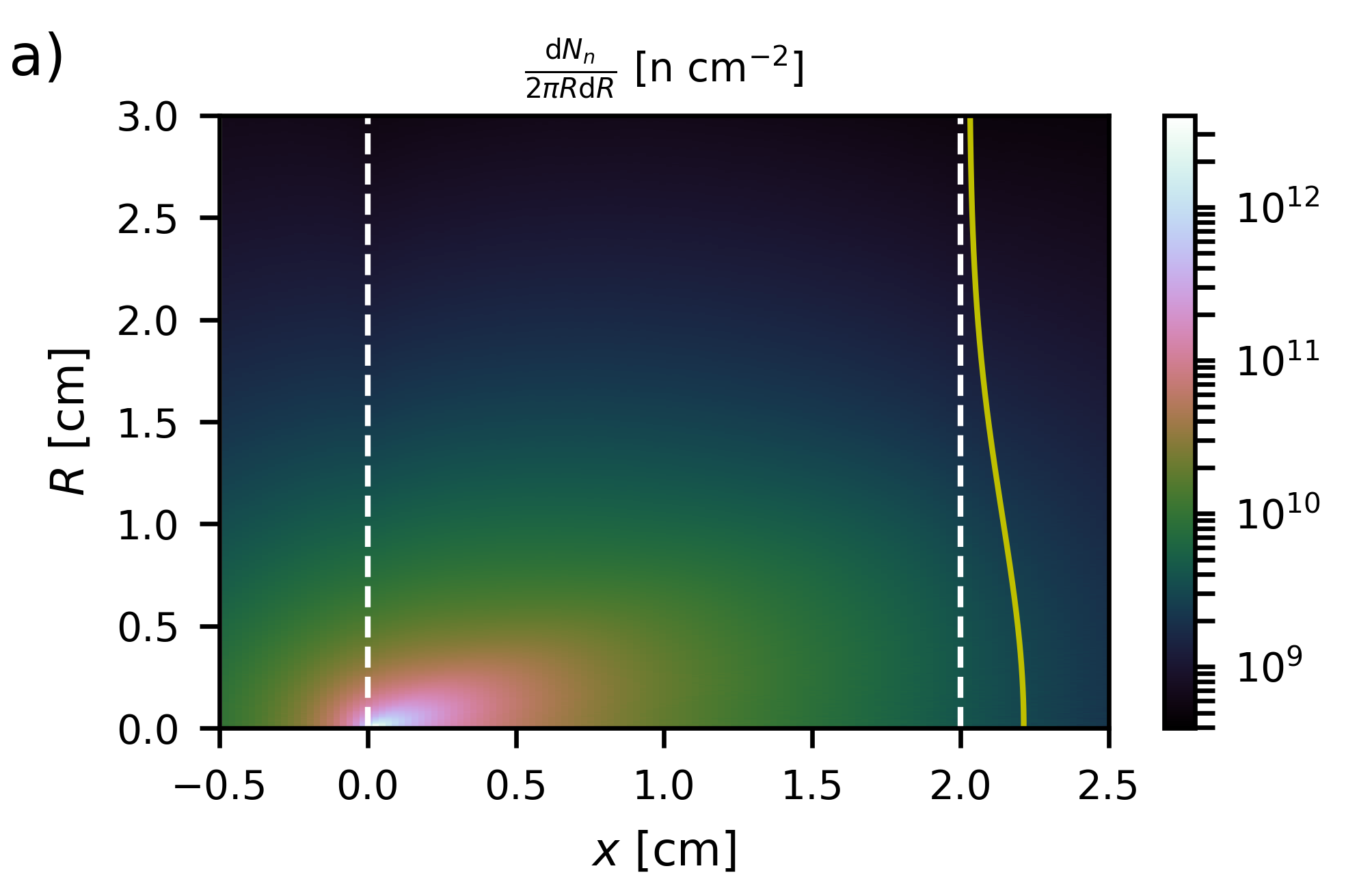}
    \includegraphics[scale=1.0]{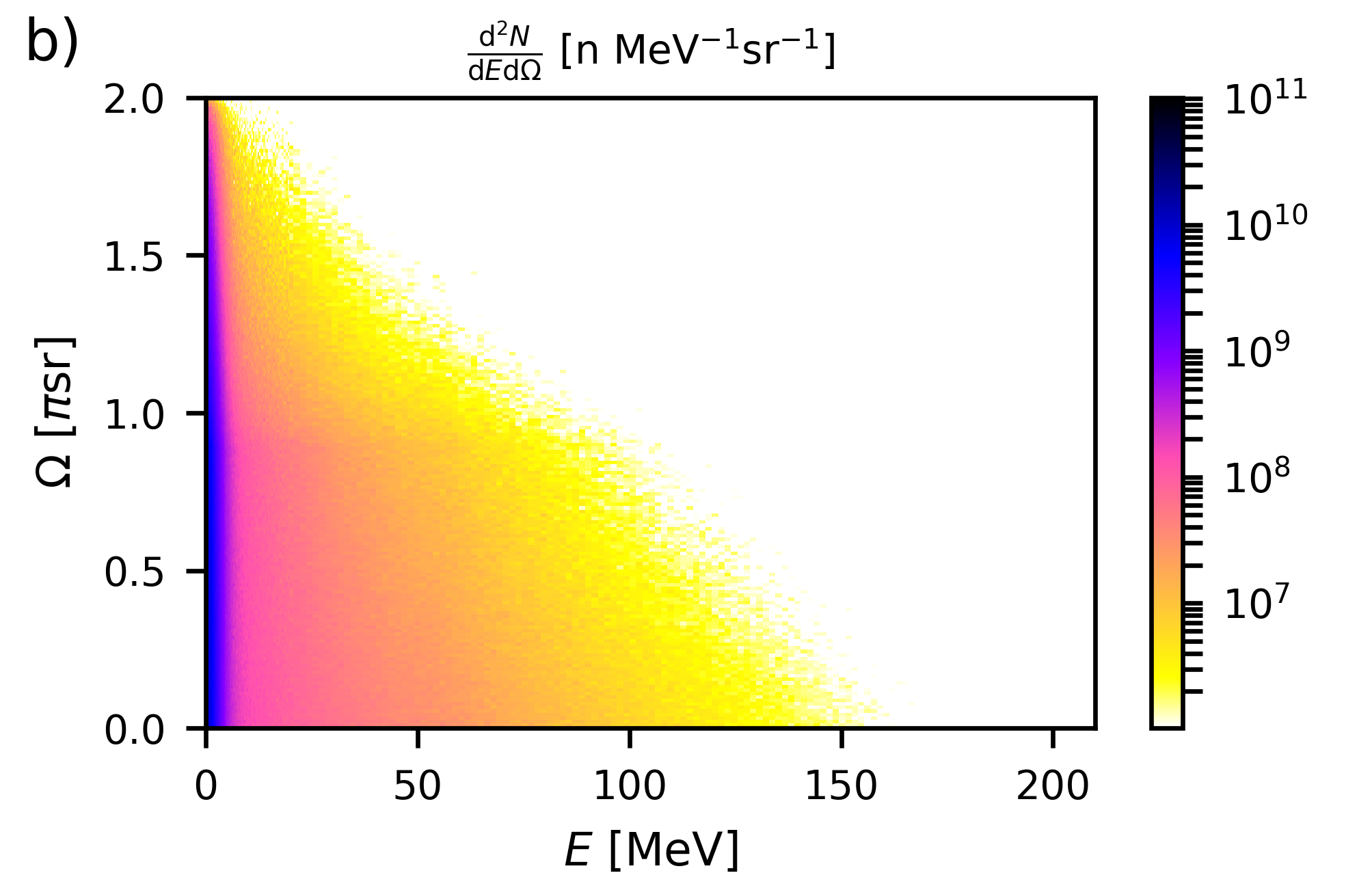}
    \caption{\textbf{Neutron fluence profile across the converter and energy-angle spectrum of the outgoing neutrons.}
    (a) Spatial distribution of the neutron fluence through a 2-cm-long lead converter. The white dashed lines indicate the front and rear sides of the converter. The primary protons, injected from the left side, are those accelerated in PIC run \#3 (see Table \ref{tab:simpar}). The yellow line is a Gaussian fit (with a FWHM width $w_n =2.3\,\rm cm$) of the neutron fluence as recorded at the converter's backside ($x=2\,\rm cm$).
    (b) Energy-angle spectrum of the outgoing neutrons. 
    }
    \label{fig:nexposure}
\end{figure}

Figure~\ref{fig:nexposure}(a) depicts the neutron fluence profile through the converter when coupling proton source \#3 with a 2-cm-thick Pb converter (i.e. the proton-based setup maximizing $N_n$ and $\Phi_n$ in Pb). The front and rear boundaries of the converter are indicated by white dashed lines. Neutron fluences exceeding $10^{12}\,\rm n\,\rm cm^{-2}$ are found at depths $x \lesssim 2\,\rm mm$ around the proton beam axis. At the backside ($x=2\,\rm cm$), the neutron fluence drops to $\sim 10^{10} \,\rm n\,cm^{-2}$ over a $\sim 2.3\,\rm cm$ FWHM transverse width. Figure~\ref{fig:nexposure}(b) shows the corresponding energy-angle neutron spectrum. As only neutrons escaping from the rear side of the converter are considered, the angular spectrum is restricted to solid angles $\le  2\pi$. Two neutron populations can be distinguished: (i) an approximately isotropic group of low-energy ($\lesssim 3\,\rm MeV$) neutrons, which amounts to $\sim 70\%$ of all generated neutrons; (ii) a group of much more energetic neutrons (up to $\sim 160\,\rm MeV$), the divergence of which decreases at higher energy.

\begin{figure*}
    \centering
    \includegraphics[scale=1.0]{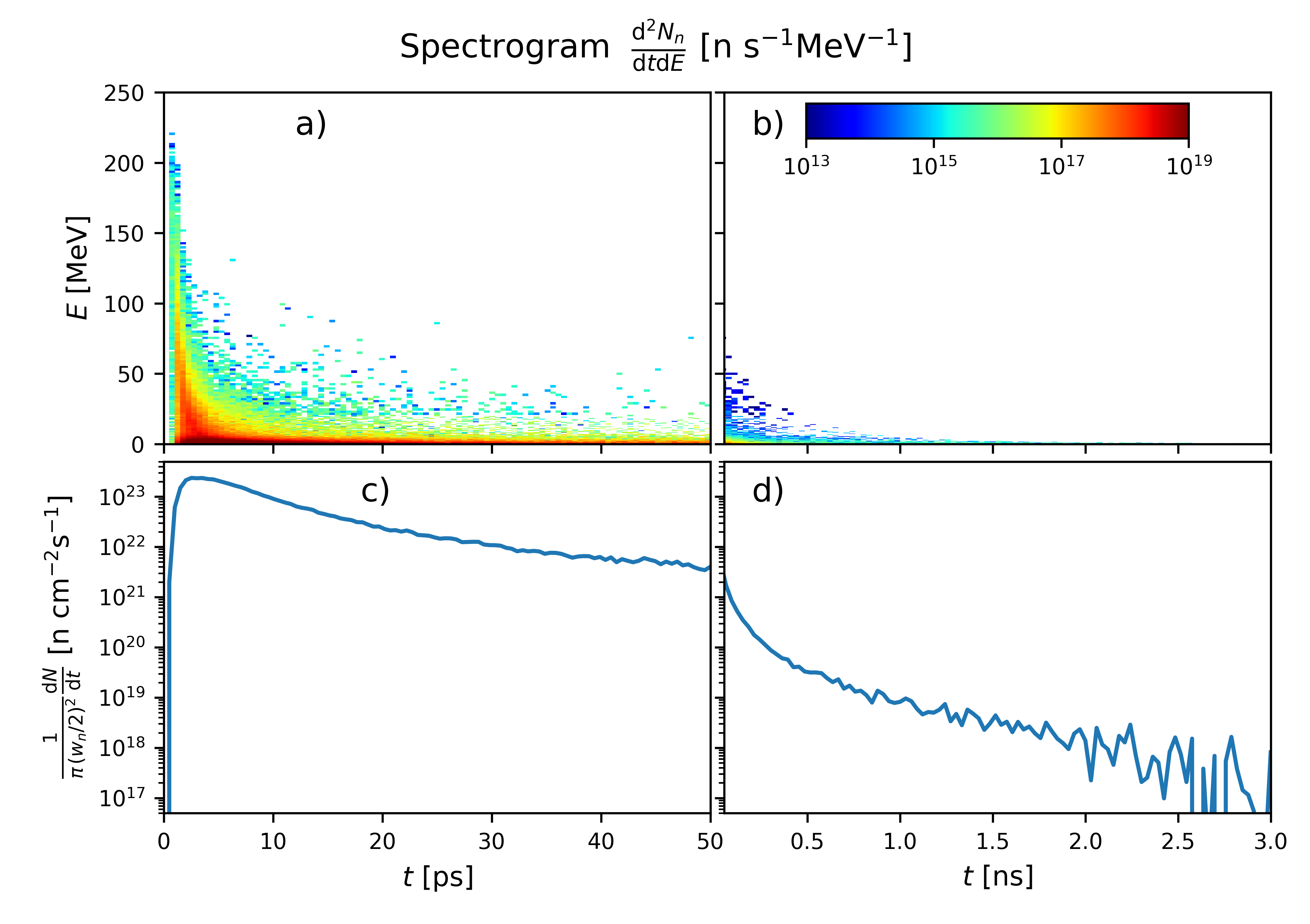}
    \caption{\textbf{Temporal variations in the neutron source.} (a,b) Time-resolved energy spectrum and (c,d) instantaneous flux of the neutrons crossing the rear side of the converter. Neutrons are here generated by sending the proton beam \#3 into a 100-$\upmu$m-thick Pb converter. Panels (a,c) detail the variations in the outgoing neutron flux over the first 50~ps while panels (b,d) show its evolution (with a cruder resolution) over a 3-ns~timespan.}
    \label{fig:nspectrogram}
\end{figure*}

Figure~\ref{fig:nspectrogram} details the temporal evolution of the neutron distribution produced from a 100-$\upmu$m-thick Pb converter by proton source \#3. This configuration generates a total number of $N_n \simeq 1.2 \times 10^9$ neutrons. The plots in the left- and right-hand side columns visualise, respectively, the short ($\le50\,\rm ps$) and long ($\le3\,\rm ns$) duration profiles of the neutron source. The early-time neutron burst, with a peak instantaneous flux of $F_n \simeq 2.3\times 10^{23}\,\rm n \,cm^{-2}\,\rm s^{-1}$, contains the most energetic neutrons (up to $\sim 230\,\rm MeV$) and is delivered within a few picoseconds (see Figs.~\ref{fig:nspectrogram}(a,b)].
The outgoing neutron flux steadily decreases afterwards, yet remains above $\sim 10^{20}\,\rm n \, cm^{-2}\, s^{-1}$ (resp. $\sim 10^{19}\,\rm n \, cm^{-2}\, s^{-1}$) till $t \simeq 270\,\rm ps$ (resp. $t \simeq 0.9\,\rm ns$). 

\section*{Discussion}

By coupling particle-in-cell and Monte Carlo simulations, we have investigated numerically the possibility of generating high-flux neutron sources by 1~PW-class, 20~fs laser pulses as are now available at the Apollon facility \cite{burdonov2021characterization}. Such sources rely on nuclear reactions triggered in a light ($^9$Be) or heavy ($^{208}$Pb) converter by fast (up to $\sim 100-200\,\rm MeV$) ions (protons or deuterons) driven from the laser target. Inspired by current trends in the laser-plasma community, we have examined the potential of double-layer targets -- comprising a few-micron-scale, near-critical plasma layer attached to a nanometric-scale solid foil -- in enhancing ion acceleration and the resulting neutron production. Our simulations predict that in specially designed DLTs, the peak intensity of the plasma-focused laser pulse can be increased nearly fourfold, entailing more than doubled maximum proton energies compared to those obtained with plain thin solid foils. This translates into approximately fourfold increased neutron yields (whether measured per unit solid angle or angle-integrated) from cm-scale Pb converters, in which $\sim  10-100\,\rm MeV$ protons benefit from large ($\sim 1-10\,\rm b$) $(p,n)$ cross sections. 
It should be noticed that besides intensifying the laser light and boosting proton acceleration, the NCD layer in DLTs can be beneficial in shielding the ultrathin solid foil against the laser pedestal or prepulses ~\cite{thaury2007plasma}.

By contrast, when employing a Be converter, more comparable neutron yields are predicted to be released by proton beams accelerated from SLTs and DLTs,  
due to decreasing $(p,n)$ cross sections at proton energies $> 5\,\rm MeV$. Interestingly, we find that the overall maximum neutron yield per unit solid angle achieved with protons (be it in Pb or Be converters) can be surpassed about twofold with deuterons by using a deuterated DLT and a Be converter.

Our simulation study also indicates that Apollon-class systems should be capable of generating peak neutron fluxes in excess of $10^{23}\,\rm n\,cm^{-2}\,s^{-1}$ using either SLTs or DLTs.
Converter targets of a few tens of microns would then be required to limit the pulse duration and lateral spread of the emitted neutrons. 

The above sources, and especially those utilising DLTs, compare favourably with previously reported experimental\cite{roth2013bright, kleinschmidt2018intense, gunther2022forward} or numerical\cite{huang2021high} works on laser-driven neutron sources but using much longer ($\tau_{\rm L} \simeq 0.5-1\,\rm ps$) and more energetic ($\sim 4-14\times$) laser pulses. This is shown in Table~\ref{tab:comparison} which summarises the results of these past works and ours, detailing in each case the laser parameters, the nature of the projectile and converter material, and the (measured or simulated) features of the neutron source.

The highest neutron yield per unit solid angle obtained in our study ($\Phi_n \simeq 1.6\times 10^{10} \,\rm n\,sr^{-1}$ in deuteron-based DLT case \#3b) is very close to the current experimental record ($\Phi_n \simeq 1.4\times 10^{10} \,\rm n\,sr^{-1}$) reported in Ref.~\cite{kleinschmidt2018intense}. Moreover, our optimum proton- and deuteron-based setups \#3 and \#3b are predicted to produce neutron yields per unit laser pulse energy of $\Phi_n/E_{\rm L} \simeq 3.4-7.0\times 10^8 \,\rm n \,sr^{-1}\,J^{-1}$, exceeding the record-high values ($\Phi_n/E_{\rm L} \simeq 0.95-2.5\times 10^8\,\rm n \,sr^{-1}\,J^{-1}$) achieved at the PHELIX facility \cite{kleinschmidt2018intense, gunther2022forward}, and performing similarly to the scheme (based on radiation pressure acceleration in overcritical CD$_2$ foams by a circularly polarised laser pulse) recently proposed in Ref.~\cite{huang2021high}.

Our numerical results appear even more promising as regards the time-averaged neutron yield per unit solid angle, that is, the product of the neutron yield and the laser shot repetition rate. This comparison evidently implies that our target setups can be adapted to the relatively high repetition rate (one shot per minute) allowed by Apollon-class systems. When combining the CH$_2$ SLT \#1 and a Pb converter of optimum length ($\sim 1\,\rm cm$), a time-averaged neutron yield of $f_{\rm L} \Phi_n \simeq 3\times 10^7\,\rm n \,sr^{-1}\,s^{-1}$ is expected, a value about ten times higher than measured experimentally so far. When attaching an optimum-density plasma layer to the optimum-thickness solid CH$_2$ layer (DLT case \#3), this value can rise to $f_{\rm L} \Phi_n \simeq 1.2\times 10^8 \,\rm n \,sr^{-1}\,s^{-1}$. Finally, using the same DLT with a solid CD$_2$ layer (case \#3b), a further increase to $f_{\rm L} \Phi_n \sim 2.6\times 10^8\,\rm n \,sr^{-1}\,s^{-1}$ is foreseen. If experimentally confirmed, this would constitute a two-orders of-magnitude improvement over the state of the art.

\begin{table}[]
\begin{tabular}{|l|rrl|c|ccc|}
\hline
Reference or simulation \#    & $E_{\rm L}$ {[}J{]} & $\tau_{\rm L}$ {[}fs{]} & $f_{\rm L}$ {[}Hz{]} & configuration   & $\Phi_n$ {[}$\rm n\,sr^{-1}${]} & $\Phi_n/E_{\rm L}$ {[}$\rm n\,J^{-1}\,sr^{-1}${]} & $f_{\rm L}\Phi_n$ {[}$\rm n\,s^{-1}\,sr^{-1}${]} \\
 \hline
\textbf{Roth \textit{et al.} (2013)}\cite{roth2013bright}                   & 80          & 600              & 2.8$\times 10^{-4}$     & d+Be         & 4.4$\times 10^{9}$             & 5.5$\times 10^{7}$                 & 1.2$\times 10^{6}$                \\
\textbf{Kleinschmidt \textit{et al.} (2018)}\cite{kleinschmidt2018intense} & 150         & 500             & 1.8$\times 10^{-4}$     & d+Be         & 1.4$\times 10^{10}$             & 9.5$\times 10^{7}$                 & 2.6$\times 10^{6}$                \\
\textbf{G\"unther \textit{et al.} (2022)}\cite{gunther2022forward} & 180         & 750             & 1.8$\times 10^{-4}$     & p+Au         & 5.2$\times 10^{9}$             & 2.7$\times 10^{8}$                 & 1.5$\times 10^{6}$                \\
\hline
Huang \textit{et al.} (2022)\cite{huang2021high}                  & 282         & 1000            & 2.8$\times 10^{-4}$     & d+Be         & 1.7$\times 10^{11}$             & 6.0$\times 10^{8}$                 & 4.7$\times 10^{7}$                \\
\hline
  \#1    & 22  & 20  & 1.7$\times 10^{-2}$  & p+Pb & 1.8$\times 10^{9}$ & 8.4$\times 10^{7}$ & 3.1$\times 10^{7}$  \\
  \#3  & 22   & 20    & 1.7$\times 10^{-2}$ & p+Pb & 7.4$\times 10^{9}$ & 3.4$\times 10^{8}$ & 1.2$\times 10^{8}$   \\
 \#3b   & 22  & 20 & 1.7$\times 10^{-2}$ & d+Be   & 1.6$\times 10^{10}$   & $7.0\times 10^{8}$  & 2.6$\times 10^{8}$  \\
\hline
\end{tabular}
\caption{\textbf{Comparison of the optimum neutron sources studied in this work with state-of-the-art published results.} The first column details the reference or simulation case considered (experimental works are highlighted in bold). The next three columns present the corresponding laser pulse energy ($E_{\rm L}$), duration ($\tau_{\rm L}$), and the shot frequency ($f_{\rm L}$). The fifth column details the ion projectile and the converter material, and the last three columns summarise the total neutron yield per unit solid angle ($\Phi_n$), the neutron yield normalised to the laser energy ($\Phi_n/E_{\rm L}$) and the time-averaged neutron yield ($f_{\rm L}\Phi_n$).}
\label{tab:comparison}
\end{table}

\section*{Methods}

\subsection*{Simulations of laser propagation in near-critical plasmas}

Laser self-focusing in near-critical-density (NCD) plasmas was investigated through 2D3V simulations (two dimensional in space, three dimensional in momentum space) performed with the PIC \textsc{calder} code\cite{lefebvre2003electron}. The simulation box had dimensions $L_x \times L_y = 63.7 \times 38.2$~$\upmu$m$^2$. It was composed of $10000 \times 6000$ cells of size $\Delta x = \Delta y = 6.37\,\rm nm$.
The laser pulse maximum entered the left side of the simulation domain at $t=40.3\,\rm fs$. A moving window technique was employed in cases where the laser pulse propagated a larger distance than the longitudinal box size. 

The NCD plasma was made of electrons and fully ionised carbon ions (C$^{6+}$). It was initialised with a uniform density profile, starting at $x=9.5\,\rm \upmu m$, and represented by 15 macroparticles per cell and species. The electron density $n_{e,\rm NCD}$ was varied from $0.45\,n_{\rm cr}$ to $2.1\,n_{\rm cr}$ over eight simulations.

\subsection*{Simulations of proton acceleration}

Ion acceleration from both SLTs and DLTs was modeled in 2D3V geometry with the PIC \textsc{calder} code. The size of the simulation domain was $L_x \times L_y = 63.7 \times 50.9$~$\upmu$m$^2$, discretised into $10000 \times 8000$ cells with $\Delta x = \Delta y = 6.37\,\rm nm$. A full 3D simulation of the problem with the same level of discretisation remains well outside our computational reach. The laser pulse maximum entered the simulation box at $t=40.32\,\rm fs$. These simulations captured both the self-focusing of the laser pulse through the NCD carbon plasma layer (if present) and its interaction with the solid-density (SD) CH$_2$ or CD$_2$ layer, from which originate the protons or deuterons used for neutron production.

The SD layer was modeled by 50 macroparticles per cell and species (electrons, C$^{6+}$, and H$^+$ or D$^+$) while 15 macroparticles per cell and species were used in the NCD layer (electrons, C$^{6+}$). Fourth-order shape factors were used for the macroparticles.

According to the PIC simulation results of Refs.~\cite{brantov2007ion, brantov2015ion}, the thickness of the SD layer that optimises the cutoff ion energy with a femtosecond laser pulse is given by
\begin{equation}
    l_{\rm opt} \simeq 0.5 a_0 \frac{n_{\rm cr}}{n_{e,\rm SD}}\lambda_{\rm L} \,,
\label{eq:brantov}
\end{equation}
where $n_{e,\rm SD}$ is the electron density of the SD layer. For the density $n_{e,\rm SD} = 200\,n_{\rm cr}$ used in our simulations, we obtain $l_{\rm opt} \simeq 62\,\rm nm$. This value served as a reference to design our SD targets.

\subsection*{Simulations of neutron generation}

The 3D Monte Carlo \textsc{fluka} code \cite{bohlen2014fluka, vlachoudis2009flair} was employed to describe the neutron generation from the fast protons reaching the virtual detector in the 2D3V \textsc{calder} PIC simulation. To this goal, we extended \textsc{fluka} with two new modules allowing the code to accept as inputs the macro-particles tracked in the PIC simulation. 

The first module enables the input particles to be injected with the same temporal profile as recorded in the PIC simulation.
The second module converts the set of macro-particles into an axisymmetric distribution. In detail, each macro-particle's position $(x, y)$ and momentum $(p_x, p_y)$ is rotated by a random azimuthal angle. Moreover, to obtain the number of physical particles represented by the macro-particle in 3D geometry, its original (2D) statistical weight (a linear density in 2D geometry) is multiplied by $2\pi y_0$, where $y_0$ is its initial distance to the axis in the unperturbed target (prior to the laser irradiation). A similar procedure was adopted by Jiang {\it et al.}.~\cite{jiang2021enhancing} to inject a 2D-PIC-simulated electron distribution into a 3D hybrid-PIC code. We note that this post-processing modifies the absolute laser-to-ion conversion efficiency: it is evaluated to be of $\sim \,9\%$ which is higher than, yet comparable with, the value of $\sim 5\%$ directly extracted from the 2D simulation. These values are consistent with the $\sim 10\,\%$ conversion efficiency reported experimentally by Higginson {\it et al.}\cite{higginson2018near} in relativistically transparent foils, albeit using a higher-energy ($\sim 200\,\rm J$), longer-duration ($\sim 0.9\,\rm ps$) laser drive.

Each \textsc{fluka} simulation made use of a minimum of $3.6 \times 10^8$ particles to model the incident proton or deuteron flux. According to convergence tests, this choice ensured the statistical accuracy of the output data. Additional simulations were also conducted to assess the contribution to neutron production of photonuclear reactions induced by fast electrons escaping the laser target. Under our conditions, this mechanism was found to be negligible compared to proton-induced neutron generation. 

As \textsc{fluka} does not yet include models for deuteron transport, we resorted to the Monte Carlo \textsc{mcnp-6} code\cite{werner2018mcnp} to simulate deuteron-induced neutron generation. These simulations did not provide output on the temporal profile of the neutron pulse. 

In all Monte Carlo simulations, the converter target was a 3-cm-radius, finite-length cylinder, made of beryllium or lead. The former material is that used in state-of-the-art laser experiments while the latter is representative of high-$Z$ converters, susceptible to spallation-type neutron generation for proton energies above $\sim 200\,\rm MeV$\cite{filges2009handbook, krasa2010spallation}.

\section*{Acknowledgments}        

V.H. acknowledges B.~Martinez for providing the modules converting the PIC \textsc{calder} simulation output to a format suitable for the Monte Carlo \textsc{fluka} code. G. Sary is also thanked for his technical assistance in the \textsc{fluka} simulations. This work was supported by funding from the European Research Council (ERC) under the European Union's Horizon 2020 research and innovation program (Grant Agreement No. 787539, project GENESIS), and by Grant ANR-17-CE30- 0026-Pinnacle from Agence Nationale de la Recherche. We acknowledge GENCI-TGCC for granting us access to the supercomputer IRENE under Grants No. A0100507594 and A0110512993.

\section*{Author contributions statement}
The overall project on bright laser-driven neutron sources has been proposed by J.F. and S.N.C., with discussion with L.G.. V.H. and L.G. set up all the simulations, except those run with \textsc{mcnp-6} which were set up by V.L.. V.H. analysed all simulations, with discussions with L.G., X.D. and J.F.. V.H. wrote the bulk of the paper, with contributions from L.G., J.F. and S.N.C..

\section*{Additional information}

\textbf{Competing financial interests} The authors declare no competing financial interests. 

\noindent\textbf{Data availability}
All data needed to evaluate the conclusions in the paper are present in the paper. Simulation results are available from the corresponding author upon reasonable request.
    


\begin{thebibliography}{10}
\urlstyle{rm}
\expandafter\ifx\csname url\endcsname\relax
  \def\url#1{\texttt{#1}}\fi
\expandafter\ifx\csname urlprefix\endcsname\relax\def\urlprefix{URL }\fi
\expandafter\ifx\csname doiprefix\endcsname\relax\def\doiprefix{DOI: }\fi
\providecommand{\bibinfo}[2]{#2}
\providecommand{\eprint}[2][]{\url{#2}}

\bibitem{takenaka1999application}
\bibinfo{author}{Takenaka, N.}, \bibinfo{author}{Asano, H.},
  \bibinfo{author}{Fujii, T.}, \bibinfo{author}{Mizubata, M.} \&
  \bibinfo{author}{Yoshii, K.}
\newblock \bibinfo{journal}{\bibinfo{title}{{Application of fast neutron
  radiography to three-dimensional visualization of steady two-phase flow in a
  rod bundle}}}.
\newblock {\emph{\JournalTitle{Nucl. Instrum. Methods Phys. Res. A}}}
  \textbf{\bibinfo{volume}{424}}, \bibinfo{pages}{73--76},
  \doiprefix\url{10.1016/S0168-9002(98)01322-9} (\bibinfo{year}{1999}).

\bibitem{mor2015reconstruction}
\bibinfo{author}{{Mor}, I.} \emph{et~al.}
\newblock \bibinfo{journal}{\bibinfo{title}{{Reconstruction of Material
  Elemental Composition Using Fast Neutron Resonance Radiography}}}.
\newblock {\emph{\JournalTitle{Phys. Procedia}}} \textbf{\bibinfo{volume}{69}},
  \bibinfo{pages}{304--313}, \doiprefix\url{10.1016/j.phpro.2015.07.043}
  (\bibinfo{year}{2015}).

\bibitem{noguere2007non}
\bibinfo{author}{Noguere, G.} \emph{et~al.}
\newblock \bibinfo{journal}{\bibinfo{title}{{Non-destructive analysis of
  materials by neutron resonance transmission}}}.
\newblock {\emph{\JournalTitle{Nucl. Instrum. Methods Phys. Res. A}}}
  \textbf{\bibinfo{volume}{575}}, \bibinfo{pages}{476--488},
  \doiprefix\url{10.1016/j.nima.2007.02.085} (\bibinfo{year}{2007}).

\bibitem{gray1943treatment}
\bibinfo{author}{Gray, L.} \& \bibinfo{author}{Read, J.}
\newblock \bibinfo{journal}{\bibinfo{title}{{Treatment of cancer by fast
  neutrons}}}.
\newblock {\emph{\JournalTitle{Nature}}} \textbf{\bibinfo{volume}{152}},
  \bibinfo{pages}{53--54}, \doiprefix\url{10.1038/152053a0}
  (\bibinfo{year}{1943}).

\bibitem{tajima2021spent}
\bibinfo{author}{Tajima, T.}, \bibinfo{author}{Necas, A.},
  \bibinfo{author}{Mourou, G.}, \bibinfo{author}{Gales, S.} \&
  \bibinfo{author}{Leroy, M.}
\newblock \bibinfo{journal}{\bibinfo{title}{Spent nuclear fuel incineration by
  fusion-driven liquid transmutator operated in real time by laser}}.
\newblock {\emph{\JournalTitle{{Fusion Sci. Technol.}}}}
  \textbf{\bibinfo{volume}{77}}, \bibinfo{pages}{251--265},
  \doiprefix\url{10.1080/15361055.2021.1889918} (\bibinfo{year}{2021}).

\bibitem{chen2019extreme}
\bibinfo{author}{Chen, S.~N.} \emph{et~al.}
\newblock \bibinfo{journal}{\bibinfo{title}{Extreme brightness laser-based
  neutron pulses as a pathway for investigating nucleosynthesis in the
  laboratory}}.
\newblock {\emph{\JournalTitle{{Matter Radiat. Extremes}}}}
  \textbf{\bibinfo{volume}{4}}, \bibinfo{pages}{054402},
  \doiprefix\url{10.1063/1.5081666} (\bibinfo{year}{2019}).

\bibitem{brown1994application}
\bibinfo{author}{Brown, D.~R.} \emph{et~al.}
\newblock \bibinfo{journal}{\bibinfo{title}{Application of pulsed fast neutrons
  analysis to cargo inspection}}.
\newblock {\emph{\JournalTitle{{Nucl. Instrum. Methods Phys. Res. A}}}}
  \textbf{\bibinfo{volume}{353}}, \bibinfo{pages}{684--688},
  \doiprefix\url{10.1117/12.171246} (\bibinfo{year}{1994}).

\bibitem{sowerby2007recent}
\bibinfo{author}{Sowerby, B.~D.} \& \bibinfo{author}{Tickner, J.~R.}
\newblock \bibinfo{journal}{\bibinfo{title}{{Recent advances in fast neutron
  radiography for cargo inspection}}}.
\newblock {\emph{\JournalTitle{{Nucl. Instrum. Methods Phys. Res. A}}}}
  \textbf{\bibinfo{volume}{580}}, \bibinfo{pages}{799--802},
  \doiprefix\url{10.1016/j.nima.2007.05.195} (\bibinfo{year}{2007}).

\bibitem{franchet1993radiolytic}
\bibinfo{author}{Franchet-Beuzit, J.}, \bibinfo{author}{Spotheim-Maurizot, M.},
  \bibinfo{author}{Sabattier, R.}, \bibinfo{author}{Blazy-Baudras, B.} \&
  \bibinfo{author}{Charlier, M.}
\newblock \bibinfo{journal}{\bibinfo{title}{{Radiolytic footprinting. Beta.
  rays, gamma photons, and fast neutrons probe DNA-protein interactions}}}.
\newblock {\emph{\JournalTitle{Biochemistry}}} \textbf{\bibinfo{volume}{32}},
  \bibinfo{pages}{2104--2110}, \doiprefix\url{10.1021/bi00059a031}
  (\bibinfo{year}{1993}).

\bibitem{constantinescu1999fast}
\bibinfo{author}{Constantinescu, B.} \emph{et~al.}
\newblock \bibinfo{journal}{\bibinfo{title}{{Fast neutron activation analysis -
  FNAA using d(13 MeV)+ Be reaction for archaeometrical research at Bucharest
  cyclotron}}}.
\newblock {\emph{\JournalTitle{{Czechoslov. J. Phys.}}}}
  \textbf{\bibinfo{volume}{49}}, \bibinfo{pages}{385--388}
  (\bibinfo{year}{1999}).

\bibitem{gratuze1992ancient}
\bibinfo{author}{Gratuze, B.}, \bibinfo{author}{Barrandon, J.-N.},
  \bibinfo{author}{Dulin, L.} \& \bibinfo{author}{Al~Isa, K.}
\newblock \bibinfo{journal}{\bibinfo{title}{Ancient glassy materials analyses:
  a new bulk nondestructive method based on fast neutron activation analysis
  with a cyclotron}}.
\newblock {\emph{\JournalTitle{{Nucl. Instrum. Methods Phys. Res. B}}}}
  \textbf{\bibinfo{volume}{71}}, \bibinfo{pages}{70--80}
  (\bibinfo{year}{1992}).

\bibitem{danson2019petawatt}
\bibinfo{author}{Danson, C.~N.} \emph{et~al.}
\newblock \bibinfo{journal}{\bibinfo{title}{Petawatt and exawatt class lasers
  worldwide}}.
\newblock {\emph{\JournalTitle{High Power Laser Sci. Eng.}}}
  \textbf{\bibinfo{volume}{7}}, \doiprefix\url{10.1017/hpl.2019.36}
  (\bibinfo{year}{2019}).

\bibitem{Henderson_2013}
\bibinfo{author}{Henderson, S.~D.}
\newblock \bibinfo{journal}{\bibinfo{title}{Spallation neutron sources and
  accelerator-driven systems}}.
\newblock {\emph{\JournalTitle{Rev. Accel. Sci. Technol.}}}
  \textbf{\bibinfo{volume}{06}}, \bibinfo{pages}{59--83},
  \doiprefix\url{10.1142/S1793626813300041} (\bibinfo{year}{2013}).

\bibitem{pomerantz2014ultrashort}
\bibinfo{author}{{Pomerantz}, I.} \emph{et~al.}
\newblock \bibinfo{journal}{\bibinfo{title}{{Ultrashort Pulsed Neutron
  Source}}}.
\newblock {\emph{\JournalTitle{Phys. Rev. Lett.}}}
  \textbf{\bibinfo{volume}{113}}, \bibinfo{pages}{184801},
  \doiprefix\url{10.1103/PhysRevLett.113.184801} (\bibinfo{year}{2014}).

\bibitem{higginson2015temporal}
\bibinfo{author}{{Higginson}, D.~P.} \emph{et~al.}
\newblock \bibinfo{journal}{\bibinfo{title}{{Temporal Narrowing of Neutrons
  Produced by High-Intensity Short-Pulse Lasers}}}.
\newblock {\emph{\JournalTitle{Phys. Rev. Lett.}}}
  \textbf{\bibinfo{volume}{115}}, \bibinfo{pages}{054802},
  \doiprefix\url{10.1103/PhysRevLett.115.054802} (\bibinfo{year}{2015}).

\bibitem{roth2013bright}
\bibinfo{author}{Roth, M.} \emph{et~al.}
\newblock \bibinfo{journal}{\bibinfo{title}{Bright laser-driven neutron source
  based on the relativistic transparency of solids}}.
\newblock {\emph{\JournalTitle{Physical Review Letters}}}
  \textbf{\bibinfo{volume}{110}}, \bibinfo{pages}{044802},
  \doiprefix\url{10.1103/PhysRevLett.110.044802} (\bibinfo{year}{2013}).

\bibitem{kleinschmidt2018intense}
\bibinfo{author}{Kleinschmidt, A.} \emph{et~al.}
\newblock \bibinfo{journal}{\bibinfo{title}{{Intense, directed neutron beams
  from a laser-driven neutron source at PHELIX}}}.
\newblock {\emph{\JournalTitle{Phys. Plasmas}}} \textbf{\bibinfo{volume}{25}},
  \bibinfo{pages}{053101}, \doiprefix\url{10.1063/1.5006613}
  (\bibinfo{year}{2018}).

\bibitem{gunther2022forward}
\bibinfo{author}{{G{\"u}nther}, M.~M.} \emph{et~al.}
\newblock \bibinfo{journal}{\bibinfo{title}{{Forward-looking insights in
  laser-generated ultra-intense {\ensuremath{\gamma}}-ray and neutron sources
  for nuclear application and science}}}.
\newblock {\emph{\JournalTitle{Nat. Commun.}}} \textbf{\bibinfo{volume}{13}},
  \bibinfo{pages}{170}, \doiprefix\url{10.1038/s41467-021-27694-7}
  (\bibinfo{year}{2022}).

\bibitem{wilks2001}
\bibinfo{author}{{Wilks}, S.~C.} \emph{et~al.}
\newblock \bibinfo{journal}{\bibinfo{title}{{Energetic proton generation in
  ultra-intense laser-solid interactions}}}.
\newblock {\emph{\JournalTitle{Phys. Plasmas}}} \textbf{\bibinfo{volume}{8}},
  \bibinfo{pages}{542--549}, \doiprefix\url{10.1063/1.1333697}
  (\bibinfo{year}{2001}).

\bibitem{mora2003plasma}
\bibinfo{author}{{Mora}, P.}
\newblock \bibinfo{journal}{\bibinfo{title}{{Plasma expansion into a vacuum}}}.
\newblock {\emph{\JournalTitle{Phys. Rev. Lett.}}}
  \textbf{\bibinfo{volume}{90}}, \bibinfo{pages}{185002},
  \doiprefix\url{10.1103/PhysRevLett.90.185002} (\bibinfo{year}{2003}).

\bibitem{yin2011three}
\bibinfo{author}{Yin, L.} \emph{et~al.}
\newblock \bibinfo{journal}{\bibinfo{title}{Three-dimensional dynamics of
  breakout afterburner ion acceleration using high-contrast short-pulse laser
  and nanoscale targets}}.
\newblock {\emph{\JournalTitle{Physical Review Letters}}}
  \textbf{\bibinfo{volume}{107}}, \bibinfo{pages}{045003},
  \doiprefix\url{10.1103/PhysRevLett.107.045003} (\bibinfo{year}{2011}).

\bibitem{jung2013boa}
\bibinfo{author}{Jung, D.}, \bibinfo{author}{Yin, L.},
  \bibinfo{author}{Albright, B.~J.} \emph{et~al.}
\newblock \bibinfo{journal}{\bibinfo{title}{Efficient carbon ion beam
  generation from laser-driven volume acceleration}}.
\newblock {\emph{\JournalTitle{New J. Phys.}}} \textbf{\bibinfo{volume}{15}},
  \bibinfo{pages}{023007}, \doiprefix\url{10.1088/1367-2630/15/2/023007}
  (\bibinfo{year}{2013}).

\bibitem{huang2021high}
\bibinfo{author}{Huang, C.-K.} \emph{et~al.}
\newblock \bibinfo{journal}{\bibinfo{title}{High-yield and high-angular-fluence
  neutron generation from deuterons accelerated by laser-driven collisionless
  shock}}.
\newblock {\emph{\JournalTitle{Appl. Phys. Lett.}}}
  \textbf{\bibinfo{volume}{120}}, \bibinfo{pages}{024102},
  \doiprefix\url{10.1063/5.0075960} (\bibinfo{year}{2022}).

\bibitem{burdonov2021characterization}
\bibinfo{author}{Burdonov, K.} \emph{et~al.}
\newblock \bibinfo{journal}{\bibinfo{title}{{Characterization and performance
  of the Apollon short-focal-area facility following its commissioning at 1 PW
  level}}}.
\newblock {\emph{\JournalTitle{{Matter Radiat. Extremes}}}}
  \textbf{\bibinfo{volume}{6}}, \bibinfo{pages}{064402},
  \doiprefix\url{10.1063/5.0065138} (\bibinfo{year}{2021}).

\bibitem{rus2017eli}
\bibinfo{author}{Rus, B.} \emph{et~al.}
\newblock \bibinfo{title}{{ELI-Beamlines: progress in development of next
  generation short-pulse laser systems}}.
\newblock In \emph{\bibinfo{booktitle}{{Research Using Extreme Light: Entering
  New Frontiers with Petawatt-Class Lasers III}}}, vol.
  \bibinfo{volume}{10241}, \bibinfo{pages}{102410J}
  (\bibinfo{organization}{International Society for Optics and Photonics},
  \bibinfo{year}{2017}).

\bibitem{gales2018extreme}
\bibinfo{author}{{Gales}, S.} \emph{et~al.}
\newblock \bibinfo{journal}{\bibinfo{title}{{The extreme light
  infrastructure{\textemdash}nuclear physics (ELI-NP) facility: new horizons in
  physics with 10 PW ultra-intense lasers and 20 MeV brilliant gamma beams}}}.
\newblock {\emph{\JournalTitle{Rep. Prog. Phys.}}}
  \textbf{\bibinfo{volume}{81}}, \bibinfo{pages}{094301},
  \doiprefix\url{10.1088/1361-6633/aacfe8} (\bibinfo{year}{2018}).

\bibitem{radier_HPLSE_2022}
\bibinfo{author}{Radier, C.} \emph{et~al.}
\newblock \bibinfo{journal}{\bibinfo{title}{{10 PW peak power femtosecond laser
  pulses at ELI-NP}}}.
\newblock {\emph{\JournalTitle{{High Power Laser Sci. Eng.}}}}
  \textbf{\bibinfo{volume}{10}}, \bibinfo{pages}{e21},
  \doiprefix\url{10.1017/hpl.2022.11} (\bibinfo{year}{2022}).

\bibitem{bayramian2008high}
\bibinfo{author}{Bayramian, A.} \emph{et~al.}
\newblock \bibinfo{journal}{\bibinfo{title}{{High-average-power femto-petawatt
  laser pumped by the Mercury laser facility}}}.
\newblock {\emph{\JournalTitle{J. Opt. Soc. Am. B}}}
  \textbf{\bibinfo{volume}{25}}, \bibinfo{pages}{B57--B61},
  \doiprefix\url{10.1364/JOSAB.25.000B57} (\bibinfo{year}{2008}).

\bibitem{levy2007double}
\bibinfo{author}{L{\'e}vy, A.} \emph{et~al.}
\newblock \bibinfo{journal}{\bibinfo{title}{Double plasma mirror for ultrahigh
  temporal contrast ultraintense laser pulses}}.
\newblock {\emph{\JournalTitle{Opt. Lett.}}} \textbf{\bibinfo{volume}{32}},
  \bibinfo{pages}{310--312}, \doiprefix\url{10.1364/OL.32.000310}
  (\bibinfo{year}{2007}).

\bibitem{Yoon_Optica_2021}
\bibinfo{author}{Yoon, J.~W.} \emph{et~al.}
\newblock \bibinfo{journal}{\bibinfo{title}{{Realization of laser intensity
  over $10^{23}\,\rm W/cm^2$}}}.
\newblock {\emph{\JournalTitle{Optica}}} \textbf{\bibinfo{volume}{8}},
  \bibinfo{pages}{630--635}, \doiprefix\url{10.1364/OPTICA.420520}
  (\bibinfo{year}{2021}).

\bibitem{filges2009handbook}
\bibinfo{author}{Filges, D.} \& \bibinfo{author}{Goldenbaum, F.}
\newblock \emph{\bibinfo{title}{Handbook of spallation research: theory,
  experiments and applications}} (\bibinfo{publisher}{Wiley},
  \bibinfo{year}{2009}).

\bibitem{martinez2022numerical}
\bibinfo{author}{Martinez, B.} \emph{et~al.}
\newblock \bibinfo{journal}{\bibinfo{title}{{Numerical investigation of
  spallation neutrons generated from petawatt-scale laser-driven proton
  beams}}}.
\newblock {\emph{\JournalTitle{{Matter Radiat. Extremes}}}}
  \textbf{\bibinfo{volume}{7}}, \bibinfo{pages}{024401},
  \doiprefix\url{10.1063/5.0060582} (\bibinfo{year}{2022}).

\bibitem{esirkepov2006laser}
\bibinfo{author}{Esirkepov, T.}, \bibinfo{author}{Yamagiwa, M.} \&
  \bibinfo{author}{Tajima, T.}
\newblock \bibinfo{journal}{\bibinfo{title}{Laser ion-acceleration scaling laws
  seen in multiparametric particle-in-cell simulations}}.
\newblock {\emph{\JournalTitle{Phys. Rev. Lett.}}}
  \textbf{\bibinfo{volume}{96}}, \bibinfo{pages}{105001},
  \doiprefix\url{10.1103/PhysRevLett.96.105001} (\bibinfo{year}{2006}).

\bibitem{brantov2015ion}
\bibinfo{author}{Brantov, A.}, \bibinfo{author}{Govras, E.},
  \bibinfo{author}{Bychenkov, V.~Y.} \& \bibinfo{author}{Rozmus, W.}
\newblock \bibinfo{journal}{\bibinfo{title}{Ion energy scaling under optimum
  conditions of laser plasma acceleration from solid density targets}}.
\newblock {\emph{\JournalTitle{{Phys. Rev. ST Accel. Beams}}}}
  \textbf{\bibinfo{volume}{18}}, \bibinfo{pages}{021301},
  \doiprefix\url{10.1103/PhysRevSTAB.18.021301} (\bibinfo{year}{2015}).

\bibitem{henig2009rpa}
\bibinfo{author}{{Henig}, A.} \emph{et~al.}
\newblock \bibinfo{journal}{\bibinfo{title}{Radiation-pressure acceleration of
  ion beams driven by circularly polarized laser pulses}}.
\newblock {\emph{\JournalTitle{Phys. Rev. Lett.}}}
  \textbf{\bibinfo{volume}{103}}, \bibinfo{pages}{245003},
  \doiprefix\url{10.1103/PhysRevLett.103.245003} (\bibinfo{year}{2009}).

\bibitem{esirkepov2004highly}
\bibinfo{author}{Esirkepov, T.}, \bibinfo{author}{Borghesi, M.},
  \bibinfo{author}{Bulanov, S.}, \bibinfo{author}{Mourou, G.} \&
  \bibinfo{author}{Tajima, T.}
\newblock \bibinfo{journal}{\bibinfo{title}{Highly efficient relativistic-ion
  generation in the laser-piston regime}}.
\newblock {\emph{\JournalTitle{Physical Review Letters}}}
  \textbf{\bibinfo{volume}{92}}, \bibinfo{pages}{175003},
  \doiprefix\url{10.1103/PhysRevLett.92.175003} (\bibinfo{year}{2004}).

\bibitem{klimo2008monoenergetic}
\bibinfo{author}{Klimo, O.}, \bibinfo{author}{Pšikal, J.},
  \bibinfo{author}{Limpouch, J.} \& \bibinfo{author}{Tikhonchuk, V.}
\newblock \bibinfo{journal}{\bibinfo{title}{Monoenergetic ion beams from
  ultrathin foils irradiated by ultrahigh-contrast circularly polarized laser
  pulses}}.
\newblock {\emph{\JournalTitle{Phys. Rev. ST Accel. Beams}}}
  \textbf{\bibinfo{volume}{11}}, \bibinfo{pages}{031301},
  \doiprefix\url{10.1103/PhysRevSTAB.11.031301} (\bibinfo{year}{2008}).

\bibitem{qiao2012dominance}
\bibinfo{author}{Qiao, B.} \emph{et~al.}
\newblock \bibinfo{journal}{\bibinfo{title}{{Dominance of radiation pressure in
  ion acceleration with linearly polarized pulses at intensities of
  $10^{21}$~W\,cm$^{-2}$}}}.
\newblock {\emph{\JournalTitle{Physical Review Letters}}}
  \textbf{\bibinfo{volume}{108}}, \bibinfo{pages}{115002},
  \doiprefix\url{10.1103/PhysRevLett.108.115002} (\bibinfo{year}{2012}).

\bibitem{gong2020proton}
\bibinfo{author}{Gong, Z.} \emph{et~al.}
\newblock \bibinfo{journal}{\bibinfo{title}{Proton sheet crossing in thin
  relativistic plasma irradiated by a femtosecond petawatt laser pulse}}.
\newblock {\emph{\JournalTitle{Phys. Rev. E}}} \textbf{\bibinfo{volume}{102}},
  \bibinfo{pages}{013207}, \doiprefix\url{10.1103/PhysRevE.102.013207}
  (\bibinfo{year}{2020}).

\bibitem{Ferri2020}
\bibinfo{author}{{Ferri}, J.}, \bibinfo{author}{{Siminos}, E.},
  \bibinfo{author}{{Gremillet}, L.} \& \bibinfo{author}{{F{\"u}l{\"o}p}, T.}
\newblock \bibinfo{journal}{\bibinfo{title}{{Effects of oblique incidence and
  colliding pulses on laser-driven proton acceleration from relativistically
  transparent ultrathin targets}}}.
\newblock {\emph{\JournalTitle{J. Plasma Phys.}}}
  \textbf{\bibinfo{volume}{86}}, \bibinfo{pages}{905860505},
  \doiprefix\url{10.1017/S0022377820000847} (\bibinfo{year}{2020}).

\bibitem{esarey1997self}
\bibinfo{author}{Esarey, E.}, \bibinfo{author}{Sprangle, P.},
  \bibinfo{author}{Krall, J.} \& \bibinfo{author}{Ting, A.}
\newblock \bibinfo{journal}{\bibinfo{title}{Self-focusing and guiding of short
  laser pulses in ionizing gases and plasmas}}.
\newblock {\emph{\JournalTitle{IEEE J. Quantum Electron.}}}
  \textbf{\bibinfo{volume}{33}}, \bibinfo{pages}{1879--1914},
  \doiprefix\url{10.1109/3.641305} (\bibinfo{year}{1997}).

\bibitem{Wang2011}
\bibinfo{author}{{Wang}, H.~Y.} \emph{et~al.}
\newblock \bibinfo{journal}{\bibinfo{title}{{Laser shaping of a relativistic
  intense, short Gaussian pulse by a plasma lens}}}.
\newblock {\emph{\JournalTitle{{Phys. Rev. Lett.}}}}
  \textbf{\bibinfo{volume}{107}}, \bibinfo{pages}{265002},
  \doiprefix\url{10.1103/PhysRevLett.107.265002} (\bibinfo{year}{2011}).

\bibitem{pazzaglia2020theoretical}
\bibinfo{author}{Pazzaglia, A.}, \bibinfo{author}{Fedeli, L.},
  \bibinfo{author}{Formenti, A.}, \bibinfo{author}{Maffini, A.} \&
  \bibinfo{author}{Passoni, M.}
\newblock \bibinfo{journal}{\bibinfo{title}{A theoretical model of laser-driven
  ion acceleration from near-critical double-layer targets}}.
\newblock {\emph{\JournalTitle{Commun. Phys.}}} \textbf{\bibinfo{volume}{3}},
  \bibinfo{pages}{1--13}, \doiprefix\url{10.1038/s42005-020-00400-7}
  (\bibinfo{year}{2020}).

\bibitem{Quesnel1998}
\bibinfo{author}{{Quesnel}, B.} \& \bibinfo{author}{{Mora}, P.}
\newblock \bibinfo{journal}{\bibinfo{title}{{Theory and simulation of the
  interaction of ultraintense laser pulses with electrons in vacuum}}}.
\newblock {\emph{\JournalTitle{{Phys. Rev. E}}}} \textbf{\bibinfo{volume}{58}},
  \bibinfo{pages}{3719--3732}, \doiprefix\url{10.1103/PhysRevE.58.3719}
  (\bibinfo{year}{1998}).

\bibitem{Salamin2002}
\bibinfo{author}{{Salamin}, Y.~I.} \& \bibinfo{author}{{Keitel}, C.~H.}
\newblock \bibinfo{journal}{\bibinfo{title}{{Electron acceleration by a tightly
  focused laser beam}}}.
\newblock {\emph{\JournalTitle{Phys. Rev. Lett.}}}
  \textbf{\bibinfo{volume}{88}}, \bibinfo{pages}{095005},
  \doiprefix\url{10.1103/PhysRevLett.88.095005} (\bibinfo{year}{2002}).

\bibitem{Pukhov1999}
\bibinfo{author}{Pukhov, A.}, \bibinfo{author}{Sheng, Z.~M.} \&
  \bibinfo{author}{Meyer-ter Vehn, J.}
\newblock \bibinfo{journal}{\bibinfo{title}{{Particle acceleration in
  relativistic laser channels}}}.
\newblock {\emph{\JournalTitle{Phys. Plasmas}}} \textbf{\bibinfo{volume}{6}},
  \bibinfo{pages}{2847--2854}, \doiprefix\url{10.1063/1.873242}
  (\bibinfo{year}{1999}).

\bibitem{arefiev2016beyond}
\bibinfo{author}{Arefiev, A.} \emph{et~al.}
\newblock \bibinfo{journal}{\bibinfo{title}{Beyond the ponderomotive limit:
  Direct laser acceleration of relativistic electrons in sub-critical
  plasmas}}.
\newblock {\emph{\JournalTitle{Phys. Plasmas}}} \textbf{\bibinfo{volume}{23}},
  \bibinfo{pages}{056704}, \doiprefix\url{10.1063/1.4946024}
  (\bibinfo{year}{2016}).

\bibitem{Debayle2017}
\bibinfo{author}{{Debayle}, A.} \emph{et~al.}
\newblock \bibinfo{journal}{\bibinfo{title}{{Electron heating by intense
  short-pulse lasers propagating through near-critical plasmas}}}.
\newblock {\emph{\JournalTitle{New J. Phys.}}} \textbf{\bibinfo{volume}{19}},
  \bibinfo{pages}{123013}, \doiprefix\url{10.1088/1367-2630/aa953f}
  (\bibinfo{year}{2017}).

\bibitem{nakamura2010foam}
\bibinfo{author}{{Nakamura}, T.}, \bibinfo{author}{{Tampo}, M.},
  \bibinfo{author}{{Kodama}, R.}, \bibinfo{author}{{Bulanov}, S.~V.} \&
  \bibinfo{author}{{Kando}, M.}
\newblock \bibinfo{journal}{\bibinfo{title}{{Interaction of high contrast laser
  pulse with foam-attached target}}}.
\newblock {\emph{\JournalTitle{Phys. Plasmas}}} \textbf{\bibinfo{volume}{17}},
  \bibinfo{pages}{113107}, \doiprefix\url{10.1063/1.3507294}
  (\bibinfo{year}{2010}).

\bibitem{sgattoni2012laser}
\bibinfo{author}{Sgattoni, A.}, \bibinfo{author}{Londrillo, P.},
  \bibinfo{author}{Macchi, A.} \& \bibinfo{author}{Passoni, M.}
\newblock \bibinfo{journal}{\bibinfo{title}{Laser ion acceleration using a
  solid target coupled with a low-density layer}}.
\newblock {\emph{\JournalTitle{Phys. Rev. E}}} \textbf{\bibinfo{volume}{85}},
  \bibinfo{pages}{036405}, \doiprefix\url{10.1103/PhysRevE.85.036405}
  (\bibinfo{year}{2012}).

\bibitem{wang2013efficient}
\bibinfo{author}{Wang, H.} \emph{et~al.}
\newblock \bibinfo{journal}{\bibinfo{title}{Efficient and stable proton
  acceleration by irradiating a two-layer target with a linearly polarized
  laser pulse}}.
\newblock {\emph{\JournalTitle{Phys. Plasmas}}} \textbf{\bibinfo{volume}{20}},
  \bibinfo{pages}{013101}, \doiprefix\url{10.1063/1.4773198}
  (\bibinfo{year}{2013}).

\bibitem{passoni2016toward}
\bibinfo{author}{Passoni, M.} \emph{et~al.}
\newblock \bibinfo{journal}{\bibinfo{title}{Toward high-energy laser-driven ion
  beams: Nanostructured double-layer targets}}.
\newblock {\emph{\JournalTitle{Phys. Rev. Accel. Beams}}}
  \textbf{\bibinfo{volume}{19}}, \bibinfo{pages}{061301},
  \doiprefix\url{10.1103/PhysRevAccelBeams.19.061301} (\bibinfo{year}{2016}).

\bibitem{levy2020enhanced}
\bibinfo{author}{Levy, D.}, \bibinfo{author}{Davoine, X.},
  \bibinfo{author}{Debayle, A.}, \bibinfo{author}{Gremillet, L.} \&
  \bibinfo{author}{Malka, V.}
\newblock \bibinfo{journal}{\bibinfo{title}{Enhanced laser-driven proton
  acceleration with gas--foil targets}}.
\newblock {\emph{\JournalTitle{J. Plasma Phys.}}}
  \textbf{\bibinfo{volume}{86}}, \doiprefix\url{10.1017/S0022377820001397}
  (\bibinfo{year}{2020}).

\bibitem{yogo2008laser}
\bibinfo{author}{Yogo, A.} \emph{et~al.}
\newblock \bibinfo{journal}{\bibinfo{title}{Laser ion acceleration via control
  of the near-critical density target}}.
\newblock {\emph{\JournalTitle{Phys. Rev. E}}} \textbf{\bibinfo{volume}{77}},
  \bibinfo{pages}{016401}, \doiprefix\url{10.1103/PhysRevE.77.016401}
  (\bibinfo{year}{2008}).

\bibitem{passoni2014energetic}
\bibinfo{author}{Passoni, M.} \emph{et~al.}
\newblock \bibinfo{journal}{\bibinfo{title}{Energetic ions at moderate laser
  intensities using foam-based multi-layered targets}}.
\newblock {\emph{\JournalTitle{Plasma Phys. Control. Fusion}}}
  \textbf{\bibinfo{volume}{56}}, \bibinfo{pages}{045001},
  \doiprefix\url{10.1088/0741-3335/56/4/045001} (\bibinfo{year}{2014}).

\bibitem{bin2015ion}
\bibinfo{author}{Bin, J.} \emph{et~al.}
\newblock \bibinfo{journal}{\bibinfo{title}{Ion acceleration using relativistic
  pulse shaping in near-critical-density plasmas}}.
\newblock {\emph{\JournalTitle{Phys. Rev. Lett.}}}
  \textbf{\bibinfo{volume}{115}}, \bibinfo{pages}{064801},
  \doiprefix\url{10.1103/PhysRevLett.115.064801} (\bibinfo{year}{2015}).

\bibitem{bin2018enhanced}
\bibinfo{author}{Bin, J.} \emph{et~al.}
\newblock \bibinfo{journal}{\bibinfo{title}{Enhanced laser-driven ion
  acceleration by superponderomotive electrons generated from
  near-critical-density plasma}}.
\newblock {\emph{\JournalTitle{Phys. Rev. Lett.}}}
  \textbf{\bibinfo{volume}{120}}, \bibinfo{pages}{074801},
  \doiprefix\url{10.1103/PhysRevLett.120.074801} (\bibinfo{year}{2018}).

\bibitem{ma2019laser}
\bibinfo{author}{Ma, W.} \emph{et~al.}
\newblock \bibinfo{journal}{\bibinfo{title}{Laser acceleration of highly
  energetic carbon ions using a double-layer target composed of slightly
  underdense plasma and ultrathin foil}}.
\newblock {\emph{\JournalTitle{Phys. Rev. Lett.}}}
  \textbf{\bibinfo{volume}{122}}, \bibinfo{pages}{014803},
  \doiprefix\url{10.1103/PhysRevLett.122.014803} (\bibinfo{year}{2019}).

\bibitem{prencipe2016development}
\bibinfo{author}{Prencipe, I.} \emph{et~al.}
\newblock \bibinfo{journal}{\bibinfo{title}{Development of foam-based layered
  targets for laser-driven ion beam production}}.
\newblock {\emph{\JournalTitle{Plasma Phys. Control. Fusion}}}
  \textbf{\bibinfo{volume}{58}}, \bibinfo{pages}{034019},
  \doiprefix\url{10.1088/0741-3335/58/3/034019} (\bibinfo{year}{2016}).

\bibitem{wang2021fabrication}
\bibinfo{author}{Wang, P.} \emph{et~al.}
\newblock \bibinfo{journal}{\bibinfo{title}{Fabrication of large-area uniform
  carbon nanotube foams as near-critical-density targets for laser--plasma
  experiments}}.
\newblock {\emph{\JournalTitle{{High Power Laser Sci. Eng.}}}}
  \textbf{\bibinfo{volume}{9}}, \doiprefix\url{10.1017/hpl.2021.18}
  (\bibinfo{year}{2021}).

\bibitem{Shadwick_PoP_2009}
\bibinfo{author}{Shadwick, B.~A.}, \bibinfo{author}{Schroeder, C.~B.} \&
  \bibinfo{author}{Esarey, E.}
\newblock \bibinfo{journal}{\bibinfo{title}{Nonlinear laser energy depletion in
  laser-plasma accelerators}}.
\newblock {\emph{\JournalTitle{Phys. Plasmas}}} \textbf{\bibinfo{volume}{16}},
  \bibinfo{pages}{056704}, \doiprefix\url{10.1063/1.3124185}
  (\bibinfo{year}{2009}).

\bibitem{brantov2007ion}
\bibinfo{author}{Brantov, A.~V.}, \bibinfo{author}{Bychenkov, V.~Y.} \&
  \bibinfo{author}{Rozmus, V.}
\newblock \bibinfo{journal}{\bibinfo{title}{Ion acceleration by ultrahigh-power
  ultrashort laser pulses}}.
\newblock {\emph{\JournalTitle{Quantum Electronics}}}
  \textbf{\bibinfo{volume}{37}}, \bibinfo{pages}{863},
  \doiprefix\url{10.1070/QE2007v037n09ABEH013486} (\bibinfo{year}{2007}).

\bibitem{babaei2017rise}
\bibinfo{author}{Babaei, J.} \emph{et~al.}
\newblock \bibinfo{journal}{\bibinfo{title}{{Rise time of proton cut-off energy
  in 2D and 3D PIC simulations}}}.
\newblock {\emph{\JournalTitle{Phys. Plasmas}}} \textbf{\bibinfo{volume}{24}},
  \bibinfo{pages}{043106}, \doiprefix\url{10.1063/1.4979901}
  (\bibinfo{year}{2017}).

\bibitem{Brown_NDS_2018}
\bibinfo{author}{Brown, D.} \emph{et~al.}
\newblock \bibinfo{journal}{\bibinfo{title}{{ENDF/B-VIII.0: The 8th Major
  Release of the Nuclear Reaction Data Library with CIELO-project Cross
  Sections, New Standards and Thermal Scattering Data}}}.
\newblock {\emph{\JournalTitle{Nucl. Data Sheets}}}
  \textbf{\bibinfo{volume}{148}}, \bibinfo{pages}{1--142},
  \doiprefix\url{https://doi.org/10.1016/j.nds.2018.02.001}
  (\bibinfo{year}{2018}).

\bibitem{Soppera_OECD_2020}
\bibinfo{author}{Soppera, N.}, \bibinfo{author}{Dupont, E.} \&
  \bibinfo{author}{Bossant, M.}
\newblock \bibinfo{title}{{JANIS book of proton induced cross sections,
  comparison of evaluated and experimental data from ENDF/B-VIII.0,
  JENDL/HE-2007, PADF-2007, TENDL-2019 and EXFOR}}.
\newblock \bibinfo{type}{Tech. Rep.}, \bibinfo{institution}{{OECD NEA Data
  Bank}} (\bibinfo{year}{2020}).

\bibitem{tain2015enhanced}
\bibinfo{author}{Tain, J.} \emph{et~al.}
\newblock \bibinfo{journal}{\bibinfo{title}{Enhanced $\gamma$-ray emission from
  neutron unbound states populated in $\beta$ decay}}.
\newblock {\emph{\JournalTitle{Phys. Rev. Lett.}}}
  \textbf{\bibinfo{volume}{115}}, \bibinfo{pages}{062502},
  \doiprefix\url{10.1103/PhysRevLett.115.062502} (\bibinfo{year}{2015}).

\bibitem{tonchev2017capture}
\bibinfo{author}{{Tonchev}, A.~P.} \emph{et~al.}
\newblock \bibinfo{journal}{\bibinfo{title}{Capture cross sections on unstable
  nuclei}}.
\newblock {\emph{\JournalTitle{Eur. J. Phys. Web Conf.}}}
  \textbf{\bibinfo{volume}{146}}, \bibinfo{pages}{01013},
  \doiprefix\url{10.1051/epjconf/201714601013} (\bibinfo{year}{2017}).

\bibitem{Chen_PoP_2014}
\bibinfo{author}{{Chen}, S.~N.} \emph{et~al.}
\newblock \bibinfo{journal}{\bibinfo{title}{{Passive tailoring of
  laser-accelerated ion beam cut-off energy by using double foil assembly}}}.
\newblock {\emph{\JournalTitle{Phys. Plasmas}}} \textbf{\bibinfo{volume}{21}},
  \bibinfo{pages}{023119}, \doiprefix\url{10.1063/1.4867181}
  (\bibinfo{year}{2014}).

\bibitem{weaver_NSE_1972}
\bibinfo{author}{Weaver, K.~A.}, \bibinfo{author}{Anderson, J.~D.},
  \bibinfo{author}{Barschall, H.~H.} \& \bibinfo{author}{Davis, J.~C.}
\newblock \bibinfo{journal}{\bibinfo{title}{{Neutron Spectra from Deuteron
  Bombardment of D, Li, Be, and C}}}.
\newblock {\emph{\JournalTitle{Nucl. Sci. Eng.}}}
  \textbf{\bibinfo{volume}{52}}, \bibinfo{pages}{35--45},
  \doiprefix\url{10.13182/NSE73-A23287} (\bibinfo{year}{1973}).

\bibitem{avrigeanu2010deuteron}
\bibinfo{author}{{Avrigeanu}, M.} \& \bibinfo{author}{{Avrigeanu}, V.}
\newblock \bibinfo{journal}{\bibinfo{title}{{Deuteron breakup effects on
  activation cross sections at low and medium energies}}}.
\newblock {\emph{\JournalTitle{{J. Phys. Conf. Ser.}}}}
  \textbf{\bibinfo{volume}{205}}, \bibinfo{pages}{012014},
  \doiprefix\url{10.1088/1742-6596/205/1/012014} (\bibinfo{year}{2010}).

\bibitem{thaury2007plasma}
\bibinfo{author}{Thaury, C.} \emph{et~al.}
\newblock \bibinfo{journal}{\bibinfo{title}{Plasma mirrors for
  ultrahigh-intensity optics}}.
\newblock {\emph{\JournalTitle{Nat. Phys.}}} \textbf{\bibinfo{volume}{3}},
  \bibinfo{pages}{424--429}, \doiprefix\url{10.1038/nphys595}
  (\bibinfo{year}{2007}).

\bibitem{lefebvre2003electron}
\bibinfo{author}{Lefebvre, E.} \emph{et~al.}
\newblock \bibinfo{journal}{\bibinfo{title}{Electron and photon production from
  relativistic laser--plasma interactions}}.
\newblock {\emph{\JournalTitle{Nucl. Fus.}}} \textbf{\bibinfo{volume}{43}},
  \bibinfo{pages}{629}, \doiprefix\url{10.1088/0029-5515/43/7/317}
  (\bibinfo{year}{2003}).

\bibitem{bohlen2014fluka}
\bibinfo{author}{B{\"o}hlen, T.} \emph{et~al.}
\newblock \bibinfo{journal}{\bibinfo{title}{{The FLUKA code: developments and
  challenges for high energy and medical applications}}}.
\newblock {\emph{\JournalTitle{Nucl. Data Sheets}}}
  \textbf{\bibinfo{volume}{120}}, \bibinfo{pages}{211--214},
  \doiprefix\url{10.1016/j.nds.2014.07.049} (\bibinfo{year}{2014}).

\bibitem{vlachoudis2009flair}
\bibinfo{author}{Vlachoudis, V.} \emph{et~al.}
\newblock \bibinfo{title}{{FLAIR: a powerful but user friendly graphical
  interface for FLUKA}}.
\newblock In \emph{\bibinfo{booktitle}{Proc. Int. Conf. on Mathematics,
  Computational Methods \& Reactor Physics (M\&C 2009), Saratoga Springs, New
  York}}, vol. \bibinfo{volume}{176} (\bibinfo{year}{2009}).

\bibitem{jiang2021enhancing}
\bibinfo{author}{Jiang, S.} \emph{et~al.}
\newblock \bibinfo{journal}{\bibinfo{title}{Enhancing positron production using
  front surface target structures}}.
\newblock {\emph{\JournalTitle{Appl. Phys. Lett.}}}
  \textbf{\bibinfo{volume}{118}}, \bibinfo{pages}{094101},
  \doiprefix\url{10.1063/5.0038222} (\bibinfo{year}{2021}).

\bibitem{higginson2018near}
\bibinfo{author}{Higginson, A.} \emph{et~al.}
\newblock \bibinfo{journal}{\bibinfo{title}{{Near-100 MeV protons via a
  laser-driven transparency-enhanced hybrid acceleration scheme}}}.
\newblock {\emph{\JournalTitle{Nat. Commun.}}} \textbf{\bibinfo{volume}{9}},
  \bibinfo{pages}{1--9}, \doiprefix\url{10.1038/s41467-018-03063-9}
  (\bibinfo{year}{2018}).

\bibitem{werner2018mcnp}
\bibinfo{author}{Werner, C.~J.} \emph{et~al.}
\newblock \bibinfo{title}{{MCNP version 6.2 release notes}}.
\newblock \bibinfo{type}{Tech. Rep.}, \bibinfo{institution}{Los Alamos National
  Lab, Los Alamos, NM (United States)} (\bibinfo{year}{2018}).

\bibitem{krasa2010spallation}
\bibinfo{author}{{Kr\'asa, A.}}
\newblock \bibinfo{title}{{Spallation reaction physics -- Neutron sources for
  ADS}}.
\newblock \bibinfo{type}{Tech. Rep.}, \bibinfo{institution}{Faculty of Nuclear
  Sciences and Physical Engineering at Czech Technical University, Prague,}
  (\bibinfo{year}{2010}).

\end{thebibliography}
\end{document}